\documentclass[twocolumn]{aastex61}

\usepackage{array,booktabs,siunitx}
\usepackage{longtable}
\usepackage{placeins}

\UseRawInputEncoding
\usepackage{float}
\usepackage{color}

\def\pers{\hbox{s$^{-1}$}}
\providecommand{\msun}{\ensuremath{\,M_\Sun}}
\providecommand{\rsun}{\ensuremath{\,R_\Sun}}
\def\coreno{\hbox{$413$}}
\providecommand{\vsini}{$v \sin i$}

\shorttitle{RVs of Nearby Mid-M Dwarfs} 
\shortauthors{Winters et al.}

\begin{document}

\title{Radial and Rotational Velocities of a Volume-Complete Sample of M Dwarfs with Masses $0.1-0.3$~M$_{\odot}$ within 15 parsecs}

\correspondingauthor{Jennifer G. Winters}
\email{j3winters@bridgew.edu}

\author[0000-0001-6031-9513]{Jennifer G.\ Winters}
\affil{Bridgewater State University, 131 Summer St., Bridgewater, MA 02325, USA}
\affil{Center for Astrophysics $\vert$ Harvard \& Smithsonian, 60 Garden Street, Cambridge, MA 02138, USA}

\author[0000-0002-1533-9029]{Emily K. Pass}
\affil{Kavli Institute for Astrophysics and Space Research, Massachusetts Institute of Technology, Cambridge, MA 02139, USA}
\affil{Center for Astrophysics $\vert$ Harvard \& Smithsonian, 60 Garden Street, Cambridge, MA 02138, USA}

\author[0000-0002-9003-484X]{David Charbonneau}
\affil{Center for Astrophysics $\vert$ Harvard \& Smithsonian, 60 Garden Street, Cambridge, MA 02138, USA}

\author{Jonathan Irwin}
\affiliation{Institute of Astronomy, University of Cambridge, Madingley Road, Cambridge CB3 0HA, UK}
\affiliation{Center for Astrophysics $\vert$ Harvard \& Smithsonian, 60 Garden Street, Cambridge, MA 02138, USA}

\author[0000-0001-9911-7388]{David W. Latham}
\affil{Center for Astrophysics $\vert$ Harvard \& Smithsonian, 60 Garden Street, Cambridge, MA 02138, USA}

\author{Jessica Mink}
\affil{Center for Astrophysics $\vert$ Harvard \& Smithsonian, 60 Garden Street, Cambridge, MA 02138, USA}

\author[0000-0002-9789-5474]{Gilbert A. Esquerdo}
\affil{Center for Astrophysics $\vert$ Harvard \& Smithsonian, 60 Garden Street, Cambridge, MA 02138, USA}

\author{Perry Berlind}
\affil{Center for Astrophysics $\vert$ Harvard \& Smithsonian, 60 Garden Street, Cambridge, MA 02138, USA}

\author{Michael L. Calkins}
\affil{Center for Astrophysics $\vert$ Harvard \& Smithsonian, 60 Garden Street, Cambridge, MA 02138, USA}

\author[0000-0001-5133-6303]{Matthew J.\ Payne}
\affil{Center for Astrophysics $\vert$ Harvard \& Smithsonian, 60 Garden Street, Cambridge, MA 02138, USA}

\begin{abstract}

We present the results from a five-year campaign to gather multi-epoch, high-resolution spectra of a volume-complete sample of 413 M dwarfs with masses 10-30\% that of the Sun that lie within 15 parsecs. We report weighted mean systemic radial velocities (RV) and rotational broadening measurements ($v \sin i$) for our targets. Our typical relative RV uncertainties are less than $50$ m \pers ~for the isolated, slowly rotating targets in our sample, and increase but remain less than 1 km \pers ~for more rapidly rotating stars. The majority of the single stars in our sample ($71\pm3$\%) have rotational broadening below our detection limit of 3.4 km \pers. When combined with astrometric data, our radial velocities allow us to calculate galactic space motions, which we use to assign thin or thick disk membership. We determine that 81\% and 8\% of our sample are highly probable thin and highly probable thick disk members, respectively. We report seven new multi-lined multiple systems and identify six additional targets with velocity variations indicative of long-period binaries, of which three are new detections. Finally, we find no significant difference in the stellar multiplicity rates of the thin disk ($22\pm2$\%) and thick disk ($21\pm8$\%) populations in our sample, implying that mid-M dwarfs are not significantly losing their companions at these relative ages. Our survey more than triples the number of these fully-convective stars with complete astrometric data and uniformly derived, multi-epoch, high-resolution RVs and rotational broadening measurements.

\end{abstract}



\section{Introduction}

Radial velocity (RV) surveys are invaluable tools in the quest to understand low-mass stellar physics. They enable the identification of the very shortest period binaries, which are not typically detectable via other methods. They also yield critical information about the activity levels of the observed stars via rotational broadening ($v \sin i$) measurements and through spectroscopic activity indicators. Furthermore, we can estimate the relative ages of the sample in question from their 3-d galactic space motions, calculated via the combination of parallax, proper motion, and RV. The availability of exquisitely precise Gaia parallaxes and proper motions mean we are no longer limited by the uncertainties on parallaxes for precise galactic space motions; but we now need precise RVs (that is, RVs with uncertainties less than 1 km \pers), which are possible only with high-resolution instruments. 

The very nearest fully-convective M dwarfs are attractive targets for exoplanet searches. High-resolution spectroscopic data are critical for exoplanet science because we will understand our exoplanet discoveries only as well as we understand their host stars. Knowledge of which stars have close stellar or brown dwarf companions is crucial for accurately measuring planetary masses and radii, and understanding the activity levels of their host stars is key to understanding the environments in which their planets were born and now reside.

RV surveys of bright M dwarfs have been highly successful in the past with low-resolution instruments, but high-resolution spectroscopic studies of fully-convective M dwarfs have been challenging due to the intrinsic faintness of even the very nearest such stars (roughly R$_{KC} = 8-16$ mag within 15 pc). The earliest RV studies of M dwarfs naturally targeted the brightest stars. \citet{Marcy(1989)} and  \citet{Tokovinin(1992b)} presented two of the first multi-epoch, high-resolution surveys of M dwarfs. \citet{Marcy(1989)} observed a sample of 72 nearby M dwarfs, of which ten were mid-M dwarfs ($0.1-0.3$ M$_{\odot}$)\footnote{While we have defined our sample based on mass limits, rather than spectral type limits, we use the term `mid-M dwarfs' or `mid-to-late M dwarfs' to refer to our sample for the remainder of this paper for simplicity}, while \citet{Tokovinin(1992b)} surveyed 200 stars of spectral types earlier than M3 V. There has been much progress in measuring RVs for nearby M dwarfs since these early efforts. But due to their low luminosities, most programs have either used mid-to-low resolution spectrographs \citep{Reid(1995), Hawley(1996), Newton(2014), Terrien(2015)}, with resulting large RV uncertainties, or have included mainly the very brightest of these red dwarfs in high-resolution spectroscopic programs \citep{Delfosse(1998), Gizis(2002), Chubak(2012), Bonfils(2013)}. Considering even lower-mass objects, some programs have sought to mitigate this low-luminosity challenge by studying the infrared spectra of the very latest M dwarfs and brown dwarfs \citep{McLean(2007),Tanner(2012), Gagliuffi(2019)}. Much progress has been made due to these studies, but more work remains to be done. A decade ago, many early-type M dwarfs had published precise RVs from high-resolution instruments, but fewer were available for the fainter and less massive, fully-convective M dwarfs (specifically, the M dwarfs with masses $0.1 - 0.3$ \msun, corresponding roughly to spectral types M3.5V - M7V).

The availability of high-resolution echelle spectrographs on 1.5-m class telescopes has provided the opportunity to spectroscopically characterize the very nearest, fully convective M dwarfs. Since September 2016, we have been conducting an all-sky, multi-epoch, high-resolution spectroscopic survey of all known M dwarfs with masses $0.1 \leq M_{\odot} \leq 0.3$ ~within 15 pc. The goals of our survey are numerous: measure multi-epoch radial (RV) and rotational velocities ($v \sin i$), measure the equivalent widths of chromospheric activity indicators, including H${\alpha}$, and identify binaries and characterize the orbits of multiple systems with periods less than 3 years. Previous papers presented spectroscopic orbits for some of the binaries and multiples in our sample \citep{Winters(2018), Winters(2020)}. Future papers from this project will present additional spectroscopic orbits, along with an analysis of the sample's stellar multiplicity characteristics. Others in our group combined our spectroscopic data with $TESS$ data on these stars, with which we measured their flare rates, H${\alpha}$ equivalent widths, and photometric rotation periods \citep{Medina(2020),Medina(2022a),Medina(2022b)}, determined the rarity of Jupiter analogs around the 200 inactive, isolated stars in the sample \citep{Pass(2023a)}, and analyzed the 123 active, effectively single stars in the sample \citep{Pass(2023b)}. Finally, we are gathering high-resolution speckle imaging data for the sample, which will complete the coverage for stellar companions at all separation regimes for this sample of nearby stars. 

Here we present RV and rotational broadening measurements for our volume-complete sample of M dwarfs with masses $0.1 \leq \msun \leq 0.3$ ~within 15 pc. From these measurements, we calculate $UVW$ space motions from which we estimate galactic population membership, and we identify new multiple systems. We describe our sample in \S \ref{sec:sample}, our data acquisition and reductions in \S \ref{sec:data}, our results in \S  \ref{sec:results}, and discuss their implications in \S \ref{sec:discussion}. We provide concluding remarks in \S \ref{sec:conclusions}.

\section{The Stellar Sample}
\label{sec:sample}

Our sample, as described in \citet{Winters(2021)}, is composed of 413 M dwarfs within 15 pc with masses $0.10 < \msun < 0.30$, of which 366 are primary stars and 47 are secondary components to more massive stars.\footnote{ We use {\it primary} to denote either a single star that is not currently known to have a companion or the most massive (or brightest in $K$) component in a multiple system; we use {\it companion} to refer to a physical member of a multiple system that is less massive (or fainter, again in $K$) than the primary star. For the eight systems that contain a white dwarf component, we consider the white dwarf to be the primary component, as it was previously the most massive component in the system.} Distances are derived from Gaia DR2 parallaxes and supplemented with ground-based parallaxes. Masses for presumed single stars, that is, stars for which no companion is known, are estimated using the mass-luminosity relation (MLR) in the $K_s$-band from \citet{Benedict(2016)}. For unresolved multiple systems, masses are either taken from the literature in cases where orbits have been measured, or deblended into their component magnitudes from which we estimate M$_K$ and then the mass for each component using the MLR. More details are given in \citet{Winters(2021)}. 

We cross-checked our sample against our internal MEarth database for existing spectra reported in the literature. These publications include those reporting radial velocities, rotational velocities, spectroscopic binary discoveries or orbit determinations. The MEarth survey target list \citep{Nutzman(2008),Irwin(2015)} was originally limited to single stars with $\mu > 0.15$\arcsec yr$^{-1}$; thus, the database typically does not include objects with low proper motions or binaries with separations $\lesssim$ 4\arcsec. For objects not included in the internal MEarth database, we used the bibliography tool in SIMBAD to search for published spectra of these objects.

We list in Table \ref{tab:lit_spec} the details of previous
large (sample sizes of more than roughly 100 targets) RV and rotational broadening studies of the stars in our sample at the time that we began our survey, separated into high- ($R > $ 19,000) and low-resolution programs. Included is the instrument used, its resolving power, and the reported RV uncertainties for each paper, as well as the number of targets surveyed and how many overlap with our sample of mid-to-late M dwarfs within 15 pc. It appears that 248 had previously measured high-resolution RVs, but many of those are of the same well-known targets. 

\begin{figure}
\hspace{-0.7cm}
\includegraphics[scale=.35,angle=0]{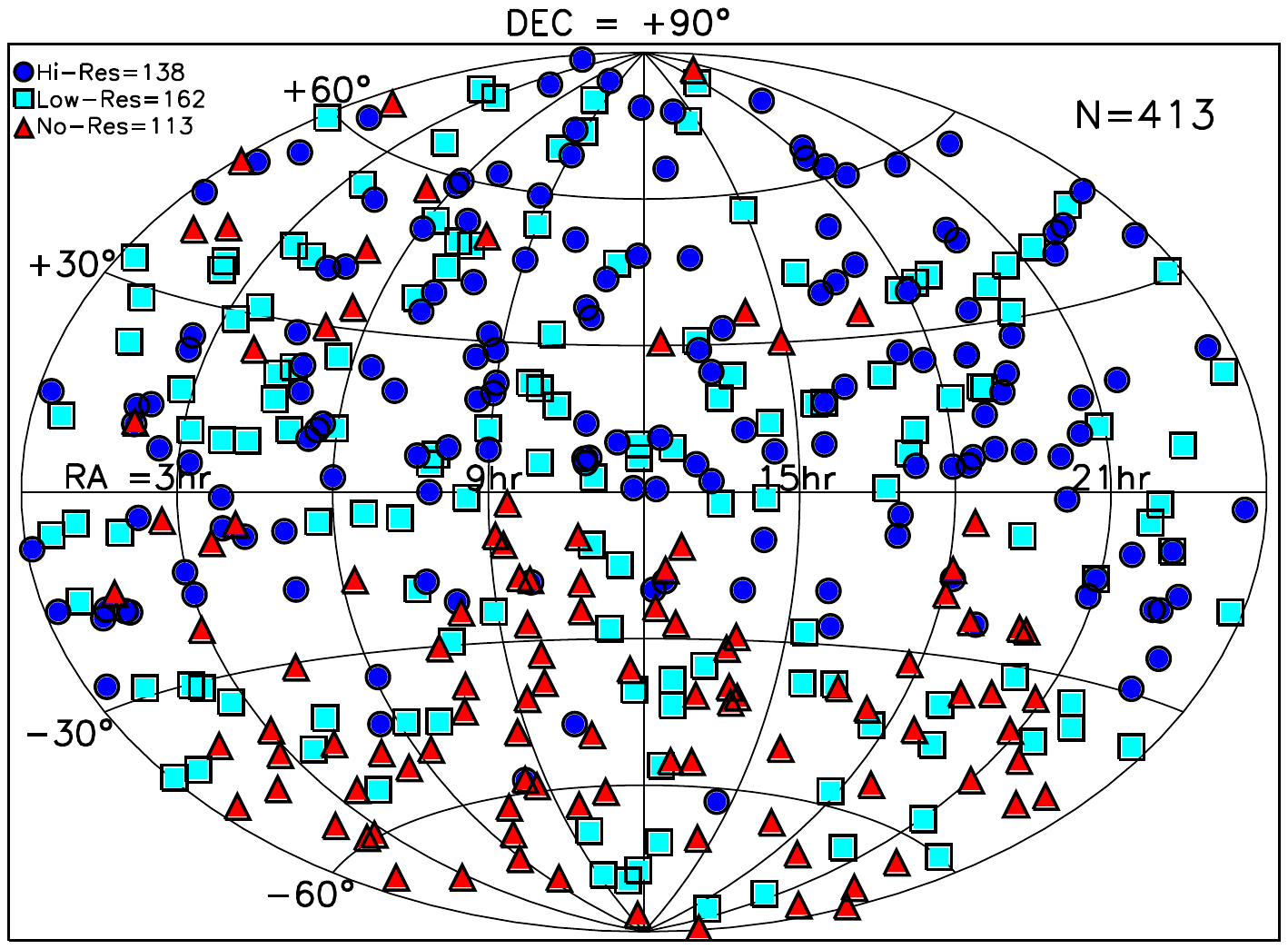}
\caption{Our targets' projected positions on the sky, in equatorial coordinates. Blue points indicate stars that had previous RVs reported in the literature from high-resolution ($R > 19,000$) instruments, cyan squares are stars with RVs measured from low-resolution instruments, while red triangles indicate stars that had no existing RV published in the literature. Of note are the large number of targets -- mostly in the southern hemisphere -- that lacked published RVs before we began our survey. \label{fig:aitoff}}
\end{figure}

We show the all-sky distribution of our spectroscopic sample of \coreno ~M dwarfs in Figure \ref{fig:aitoff} in equatorial coordinates. Of note are the large number of southern stars in our sample for which no RV data were published before we began our survey.

We note that results from the Gaia DR3 \citep{GaiaDR3(2023), Katz(2023)}, the CARMENES program \citep{Quirrenbach(2020),Reiners(2018), Jeffers(2018), Lafarga(2020)}, the SPIROU program \citep{Donati(2020), Fouque(2018)}, and another large $v \sin i$ survey \citep{Kesseli(2018)} were published while our survey was in progress. We compare our rotational broadening and RV measurements to the results of some of those surveys in \S \ref{subsubsec:vsini_msmnts} and \S \ref{subsubsec:rv_msmts}.

\begin{deluxetable*}{lccccc}
\tabletypesize{\small}
\tablecaption{Previous Large RV \& Rotational Broadening Surveys \label{tab:lit_spec}}
\tablecolumns{6}
\tablewidth{0pt}
\tablehead{\colhead{Paper}        &
    \colhead{\# Targets}          &
    \colhead{\# Overlap}           &
	\colhead{Instrument}          &
    \colhead{Resolving Power}          &
	\colhead{Uncertainty}         
}

\startdata
Radial Velocity Surveys &&&&& \\
\hline
& High-Resolution Surveys &&&& \\
\hline
\citet{Bonfils(2013)}   &  102  & 44   &  HARPS   &    115000 &   1-3 m \pers \\
\citet{Chubak(2012)}    & 2046  & 18  &   HIRES   &    55000 &    100 m \pers   \\ %
\citet{Delfosse(1998)}  &   99  & 43  &   ELODIE  &    42000 &   $15 - 70$ m s$^{-1}$ \\ 
\citet{Gizis(2002)}     & 676   & 134 &  Palomar 60-in echelle    &    19000 &    1.5 km \pers   \\ %
\citet{Shkolnik(2012)}  & 165   & 9   &  HIRES, ESPaDOnS & $>$58000 & $<$1 km \pers      \\ 
\hline
& Low-Resolution Surveys &&&& \\
\hline
\citet{Hawley(1996)}    & 1971  & 248 &  CTIO 1.5-m Cassegrain &   2000  &    $10-15$ km \pers   \\ 
\citet{Newton(2014)}    & 447   & 114 &   SpeX    &     2000 &    4 km \pers       \\ %
\citet{Reid(1995)}      & 1746 & 211 & Palomar 60-in echelle* & 2000 & $10-15$ km \pers \\
\citet{Terrien(2015)}   & 886   & 133 &   SpeX    &     2000 &   20 km \pers       \\ %
\citet{West(2015)}      & 238   & 55  &   FAST   &     3000 &    5.1 km \pers     \\ %
\hline
Rotational Velocity Surveys &&&&& \\
\hline
\citet{Browning(2010)}  & 123   & 16 &   HIRES   &    45000 &  \nodata \\ 
\citet{Delfosse(1998)}  & 101   & 43 &   ELODIE  &    42000 &  \nodata \\
\citet{Deshpande(2013)} &  253  & 5  &  APOGEE   &    22500 &  \nodata  \\ %
\enddata


\tablenotetext{*}{Forty-nine faint targets were observed with the double spectrograph on the Hale 100-in telescope. }
\end{deluxetable*}

\section{Observations and Analysis} \label{sec:data}

\subsection{Data Acquisition}
\label{subsec:acquisition}

\begin{figure}
\hspace{-0.5cm}
\includegraphics[scale=.42,angle=0]{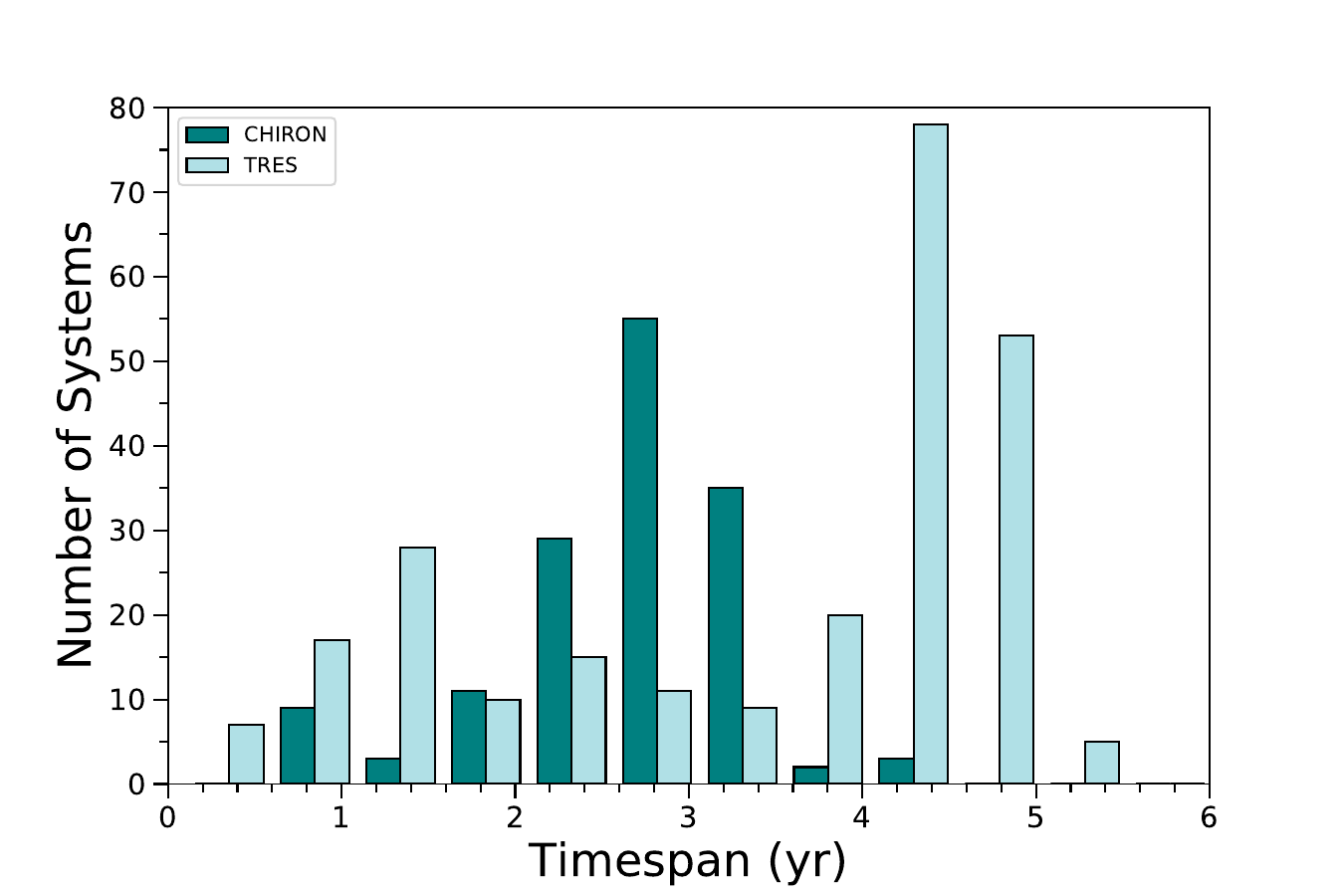}
\caption{Histogram of the largest baseline for each target. The majority of our observations were taken over five years. Not shown are baselines for 11 TRES targets which span 6-12 yrs.   \label{fig:deltat_hist}}
\end{figure}

We observed all but five of the 413 stars in our core sample at least four times between UT 2016 June 18 and 2023 July 01 using the Tillinghast Reflector Echelle Spectrograph (TRES; $R \approx 44,000$) on the 1.5m telescope at the Fred Lawrence Whipple Observatory (FLWO) on Mt. Hopkins, AZ, ~or with CTIO HIgh ResolutiON (CHIRON; $R \approx 80,000$ via slicer mode) spectrograph at the Cerro Tololo Inter-American Observatory / Small and Moderate Aperture Research Telescope System (CTIO / SMARTS) 1.5m telescope for targets roughly south of $\delta = -15$\arcdeg. GJ~334B and WIS~J1824 have only two observations each because they were added to the sample late. We observed each target over at least six months. Some of the stars in our sample had existing spectra in the TRES archive that were taken via other scientists' programs, which we included in our dataset. We show in Figure \ref{fig:deltat_hist} the largest baseline acquired for each target in our sample with timespans up to six years. The majority of our observations were taken over five years, although a handful were compiled from the internal TRES archive. Eleven additional observations of TRES targets with timespans 6-12 yrs are omitted from Figure \ref{fig:deltat_hist}.

TRES is a high-throughput, cross-dispersed, fiber-fed, echelle spectrograph. We used the medium fiber ($2\farcs3$ diameter) for a resolving power of $R \simeq 44\,000$. The spectral resolution of the instrumental profile is 6.7 km s$^{-1}$ at the center of all echelle apertures. For calibration purposes, we acquired a thorium-argon hollow-cathode lamp spectrum through the science fiber both before and after every science spectrum. Exposure times for the TRES observations ranged from $120$s to $3\times1200$s in good conditions, achieving a signal-to-noise ratio of 3-25 per pixel at $7150\ {\rm \AA}$ (the pixel scale at this wavelength is $0.059\ {\rm \AA ~pix^{-1}}$). These exposure times were increased where necessary in poor conditions. The spectra were extracted and processed using the standard TRES pipeline \citep{Buchhave(2010)}.

CHIRON is a high-throughput, cross-dispersed, fiber-fed, echelle spectrograph \citep{Tokovinin(2013)}. We used the slicer mode ($2\farcs3$ diameter) for a resolving power of $R \simeq 80\,000$. The spectral resolution of the instrumental profile is 3.8 km s$^{-1}$ at the center of all echelle apertures. As with TRES, we acquired a thorium-argon hollow-cathode lamp spectrum through the science fiber before and after every science spectrum for calibration purposes. Exposure times for the CHIRON observations ranged from $120$s to $3\times1800$s in good conditions, achieving a signal-to-noise ratio of 3-25 per pixel at $7150\ {\rm \AA}$ (the pixel scale at this wavelength is $0.025\ {\rm \AA ~pix^{-1}}$). As with the TRES observations, these exposure times were increased where necessary in poor conditions. The spectra were extracted and processed using a modified pipeline \citep{Paredes(2021)} that is based on the original by \citet{Tokovinin(2013)}.

As noted in \citet{Winters(2021)}, we obtained spectroscopic data for an additional 120 stars that were eventually discarded from the final sample. These objects were observed at least once to determine if their spectra were multi-lined, the argument being that had they been binaries, the individual masses may have moved them into our sample. In an additional seven cases, we observed the more massive M dwarf in a widely-separated binary system. We measured RVs for these objects, as well, and include them in Tables \ref{tab:rv-data} and \ref{tab:sample_data} for completeness. Except for GJ~865, which is a candidate member of our core spectroscopic sample, all 127 extra targets are marked with the `n' code to differentiate them from our core sample. GJ~865 is noted with code `c'.\\

\subsection{Rotational and Radial Velocity Measurements} \label{sec:rv_analysis}

The primary goal of our program was to measure multi-epoch RVs of our targets. However, an accurate $v \sin i$ is required to properly calculate the RV of each target. We first describe our rotational velocity measurements, followed by a description of our radial velocity measurements and their uncertainties. 

\subsubsection{Rotational Velocities}
\label{subsubsec:vsini_msmnts}

The majority of our sample (328 targets) is composed of effectively single stars, that is, either presumed single stars (stars for which no companion has been detected) or stars that are part of multiple systems with angular separation $\rho >$4\arcsec ~from its companion. Rotational broadening ($v \sin i$) measurements are straight-forward for these targets.  

We first use our original method, as described in \citet{Winters(2018)}, to measure the $v \sin i$ and estimate the RV of the target in question, using Barnard's Star as a template and utilizing only one echelle aperture centered on 7115 \AA. For small $v \sin i$, we convolve the rotational broadening kernel in the resampled (in log $\lambda$) spectra used for the CCF calculation, which are oversampled by a factor of 32 compared to the input spectrum. In cases where there is measurable rotational broadening, we measure the $v \sin i$ per target spectrum as follows. We perform two nested grid searches for the maximum peak correlation over a $v \sin i$ range of $0-100$ km s$^{-1}$ sampled at 1 km s$^{-1}$, followed by $\pm$ 1 km s$^{-1}$ about the best of those sampled at 0.1 km s$^{-1}$. We use parabolic interpolation to determine the final value of $v \sin i$ from the 0.1 km s$^{-1}$ grid results. We then average the $v \sin i$ over all the spectra gathered per object and use the average to derive the RVs, as described below in \S \ref{subsubsec:rv_msmts}. We consider only $v \sin i$ measurements larger than half the spectral resolution of each spectrograph to be robust, that is, rotational broadening larger than $3.4$ km \pers ~and $1.8$ km \pers ~for TRES and CHIRON, respectively.

While few of our sample targets had published $v \sin i$ measurements when we began our study, as illustrated in Table \ref{tab:lit_spec}, three notable studies of nearby M dwarfs using high-resolution spectrographs were published while our program was underway. Of the 328 M dwarfs from the CARMENES program \citep{Quirrenbach(2020)}\footnote{\url{https://carmenes.caha.es/ext/conferences/CARMENES_SPIE2020_Quirrenbach.pdf}} described in \citet{Reiners(2018)}, 86 targets overlap with our sample. \citet{Fouque(2018)} reported $v \sin i$ measurements from Spirou \citep{Donati(2020)} for 440 nearby M dwarfs, of which 66 overlap with our sample stars. Out of 88 M dwarfs, \citet{Kesseli(2018)} reported $v \sin i$ measurements for 30 of the stars in our sample using mainly the  Immersion Grating INfrared Spectrometer (IGRINS) on the Lowell Discovery Telescope (LDT, formerly known as the Discovery Channel Telescope) and iSHELL on the InFrared Telescope Facility. We note that the majority of the targets in these three works are northern M dwarfs.

Combining the numbers of objects with literature $v \sin i$ values from Table \ref{tab:lit_spec} with those from more recent results mentioned above \citep{Reiners(2018),Fouque(2018),Kesseli(2018)} implies that there are 246 measurements that overlap with our sample, but many surveys observed the same well-known stars. We compare only effectively single stars with detectably broadened measurements, as those are easiest to measure, which results in 55 unique comparison stars. While our detection limit for measurements with CHIRON is 1.9 km \pers, for simplicity, we adopt the TRES limit of 3.4 km \pers ~as our detection limit for the subsequent discussion. We note that rotational broadening measurements are available from Gaia from the Radial Velocity Spectrometer, which has a median resolving power 11,500, for stars with T$_{eff} > 3500$ K \citep{Fremat(2023)}. We find 27 of our stars have such measurements; however, these measurements are from moderate resolution spectra and thus have large reported uncertainties (20\% - 97\%). As with other lower-resolution $v \sin i$ measurements from the literature, we do not include these in any of our comparison analyses.  

\begin{figure}[h]
\includegraphics[scale=.39, angle=180]{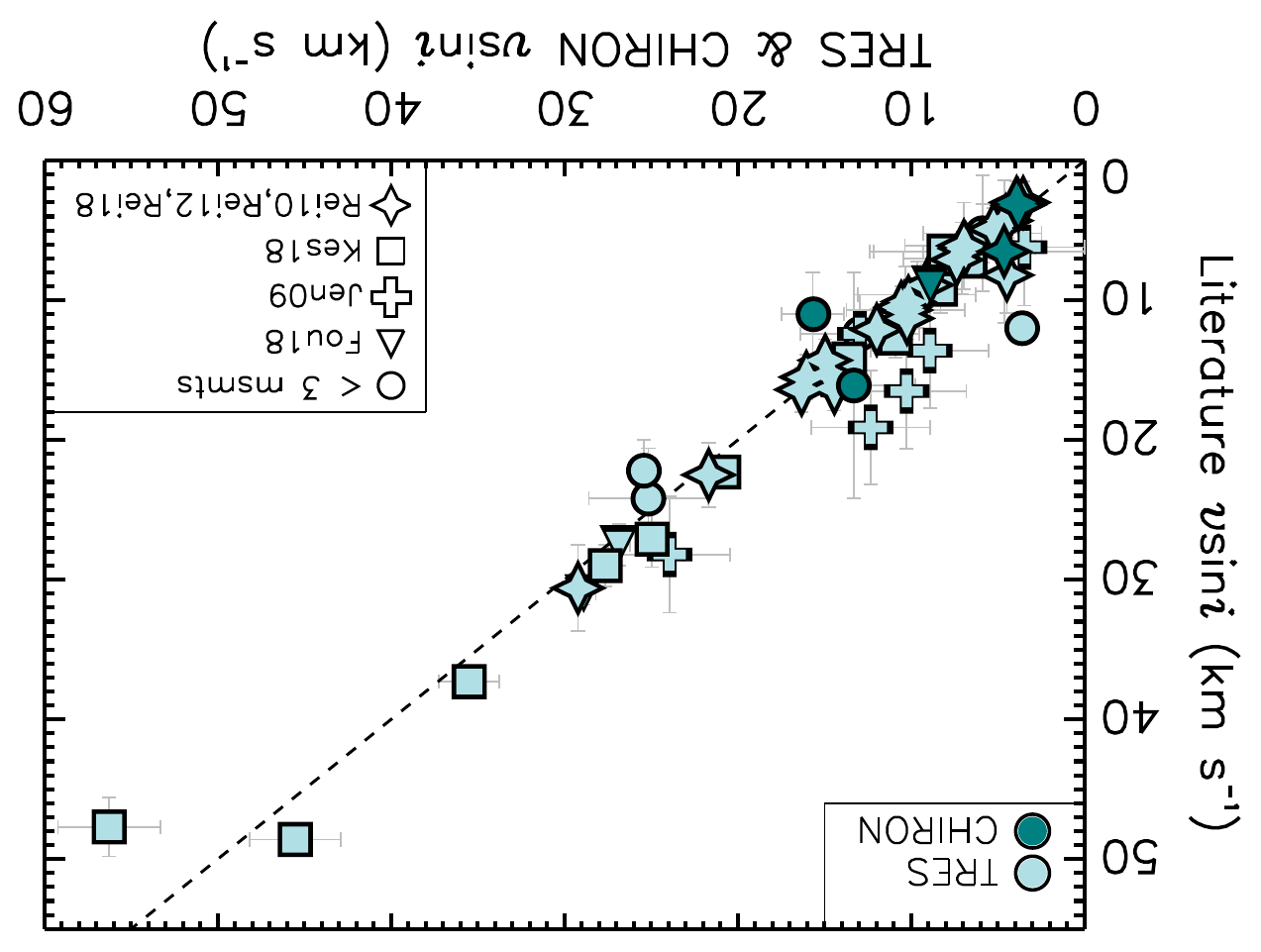}
\caption{Comparison of literature high-resolution $v \sin i$ measurements to ours for 55 effectively single objects with rotational broadening larger than our detection limit of 3.4 km \pers. TRES measurements are shown in pale blue; CHIRON measurements shown in teal. Publications presenting $v \sin i$ measurements for fewer than three distinct targets for comparison are shown as circles \citep{Barnes(2014),Browning(2010),Delfosse(1998),Deshpande(2012),Jeffers(2018),Mamajek(2013),Riedel(2011)}. In general, we see good agreement between our values and previously published values. (References: \citealt{Fouque(2018),  Jenkins(2009), Kesseli(2018), Reiners(2010), Reiners(2012), Reiners(2018)}.)  \label{fig:vsini_comp}}
\end{figure}

We illustrate in Figure \ref{fig:vsini_comp} a comparison of our measured $v \sin i$ values for the 55 unique stars with high-resolution measurements in the literature with measurements larger than 3.4 km \pers. The uncertainties on our measurements shown in Figure \ref{fig:vsini_comp} reflect half the spectral resolution of each instrument added in quadrature to the standard deviation in the four $v \sin i$ measurements for each target. In cases where no uncertainties were reported on published $v \sin i$ values (all here except \citet{Mamajek(2013)}, \citet{Fouque(2018)}, \citet{Kesseli(2018)}, and \citet{Reiners(2018)}), we have adopted uncertainties of half the spectral resolution of each spectrograph. 

As illustrated in Figure  \ref{fig:vsini_comp}, our measurements are in agreement with most measurements, although a few outliers are seen. The 24 measurements in common between our sample and measurements by Reiners \citep{Reiners(2010),Reiners(2012),Reiners(2018)} range in difference from $-3.7$ km \pers ~to $0.9$ km \pers, with a median difference of $0.08$ km \pers. When comparing our measurements to those by \citet{Kesseli(2018)}, we calculate a range of differences from $-2.8$ km \pers ~to $8.6$ km \pers, with a median difference of $-1.4$ km \pers. We note that all of the five measurements from \citet{Jenkins(2009)} are noticeably larger than our values with a range of $-6.8$ km \pers ~to $-2.7$ km \pers and a median offset of $-4.7$ km \pers. Finally, the three measurements in common with those from \citet{Fouque(2018)} have a difference range of $-2.1$ km \pers ~to $-0.01$ km \pers, with a median difference of $-0.6$ km \pers. We note that largest outlier in our comparison to \citet{Kesseli(2018)} is LSPM~J0330+5413 ($47.7 \pm 2.1$ km \pers), which differs from our median value of $56.26 \pm 2.96$ by more than $3\textrm{--}\sigma$. This system may be a binary with a composite spectrum, as our measured $v \sin i$ increases steadily over the four epochs (51.5, 55.6, 56.9, 58.4 km \pers) over which our data were taken. The data for the value from \citet{Kesseli(2018)}  were taken between our first and second observation and is marginally in agreement with our earlier measurements. In comparison, LSPM~J0357+4107 is even fainter (with the same mass), but has good agreement with our measurement ($\Delta v \sin i$ of 1.6 km \pers), so it is unlikely that the disagreement for LSPM~J0330+5413 is due to template mismatch.

A slight systematic offset in seen Figure \ref{fig:vsini_comp}, especially for stars with rotational broadening above 20 km \pers. This could be due to a template mismatch because of our use of Barnard's Star as the template for our entire sample. Barnard's Star has R$_{KC} = 8.35$ mag, mass $0.155\pm0.014$ \msun, and an undetected $v \sin i$.

For rapid rotation, any offset should be dominated by
our rotational broadening kernel. We investigated this possibility and find that for the lowest mass, rapidly rotating stars, we see up to a 4\% (2.0 km \pers ~offset for $v \sin i$ of 50 km \pers) \textit{under}estimate of rotational broadening for the very lowest mass targets, due to limb darkening effects. Our Barnard's Star template includes the use of fixed quadratic limb darkening coefficients  (0.4629, 0.3212) from PHOENIX model atmospheres for a star with T$_{eff} = 3224$K and log$_{10} g=5$ in the R$_{c}$ band from \citet{Claret(2012)}. This underestimate could be part of what we see in Figure \ref{fig:vsini_comp} for the most rapidly rotating ($v \sin i >$ 20 km \pers) stars. However, most of the comparison points in that part of plot are from \citet{Kesseli(2018)}, which used a linear limb darkening treatment and estimated rotation from the cross-correlation function FWHM using a calibration curve to relate it to rotational broadening. This different limb-darkening treatment likely also plays a role in the apparent offset seen in Figure \ref{fig:vsini_comp} at large rotational broadening values. We find no offset for the hotter (more massive targets) due to this fixed quadratic limb darkening coefficient treatment. The 12 lowest-mass, rapidly rotating ($v \sin i >$ 20 km \pers) targets in our sample would be affected by this, but the 
$v \sin i$ rms for these targets ranges from 0.3 km \pers ~to 10 km \pers. So, we consider that this limb darkening effect is already taken into account, and we perform no correction to our measurements.

We note that we previously explored the possibility of any offset in \citet{Pass(2023b)} by comparing our $v \sin i$ measurements to $v$ derived from photometric rotation period and stellar radius. We found that the $v \sin i$ distribution was in quite good agreement with expectations from isotropy, with a (marginally statistically significant) lack of low $\sin i$ stars that may be due to it being more difficult to measure rotation periods for pole-on stars.

We also explore the effect of template mismatch on our $v \sin i$ results for slowly rotating stars, where we might expect the largest effect. We tested 
coverage of the relevant parameter space using PHOENIX stellar models \citep{Husser(2013)}. We varied the effective temperature by $\pm$ 500 K and
measured the models relative to $T_{eff} =$ 3200~K (roughly Barnard's Star effective temperature) to approximate our analysis process. This was done at the resolution and wavelength range of the TRES echelle order 41, but without simulating telluric lines. We left rotational broadening as a free parameter in this analysis, as in the usual analysis of our spectroscopic data. We see zero offset in rotational broadening for stars with effective temperatures of 3200~K and hotter, but the offset increases with cooler temperatures to roughly 2.3 km \pers ~at 2700 K. Due to the lower mass limit of 0.1 \msun ~for stars in our sample, we should not have targets of spectral types M8 or M9 (effective temperatures below roughly 2600~K) in the sample, which would be affected by an even larger systematic error in rotational broadening. This analysis thus indicates that we see an \textit{over}estimate of up to 2.3 km \pers ~for our $v \sin i$ measurements for the single, slowly rotating, lowest-mass targets. We have 78 such targets in our sample that would be affected. However, we report only upper limits of 1.8 km \pers ~and 3.4 km \pers ~for these slowly-rotating stars, so we consider that this is offset is already taken into account. We perform no correction to our measurements.

We note that low-mass, late-M stars are typically faint
and rapidly rotating, which means makes it challenging to identify a bright, slowly rotating example to use as an observed template. Examination of our sample indicates that this is indeed the case. The two next brightest and slowest rotating candidates are GJ~905 (R$_{KC} = 10.77$ mag, mass $0.140\pm0.014$ \msun, $v \sin i <$ 1.0 km \pers) and GJ~54.1 (R$_{KC} = 10.73$ mag, mass $0.133\pm0.014$ \msun, $v \sin i =$ 1.48 km \pers). Neither of these are a significant improvement over our chosen template star Barnard's Star.

\begin{figure}
\includegraphics[scale=.4,angle=0]{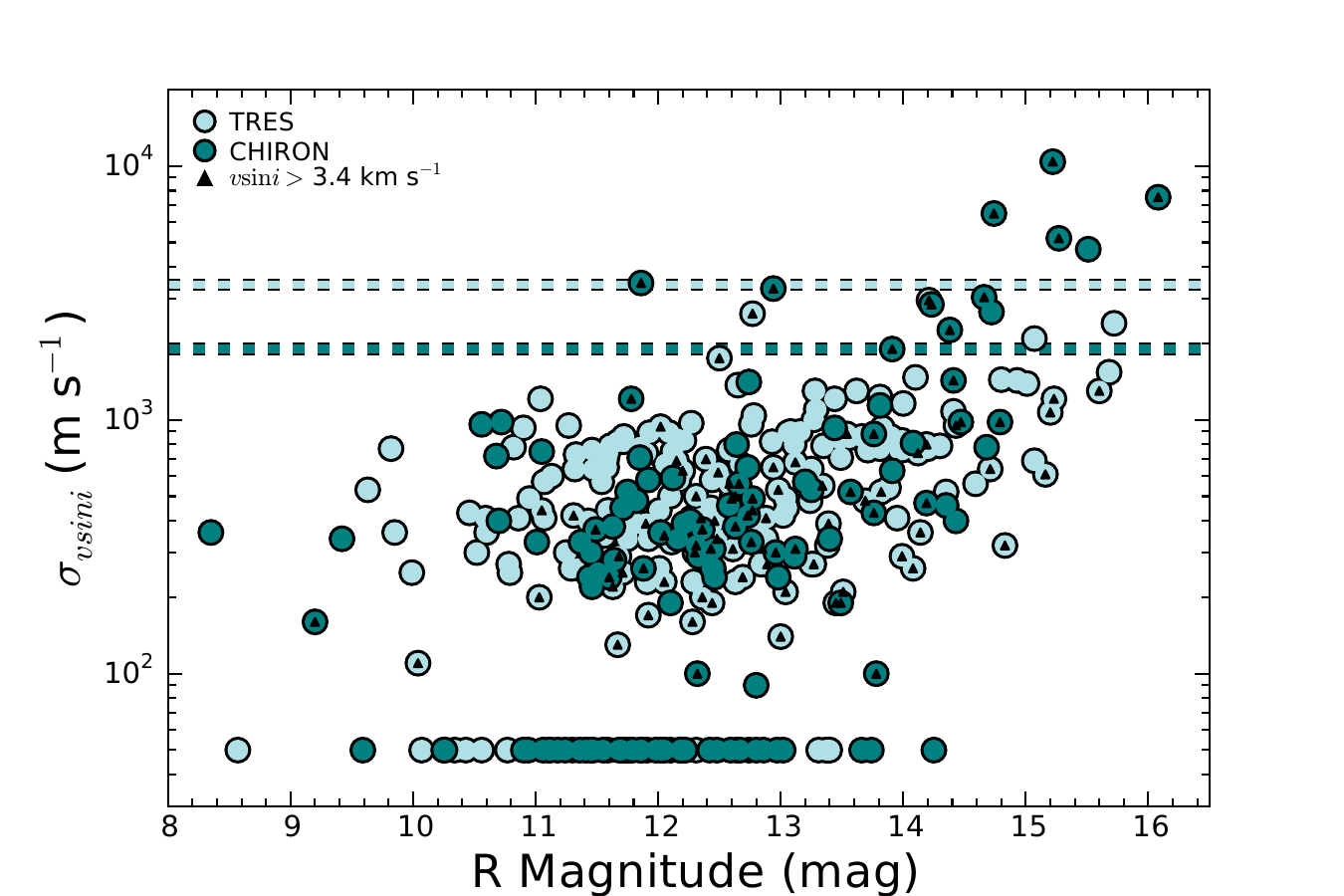}
\caption{Standard deviation (rms) of rotational broadening as a function of $R-$band magnitude for the 328 presumed single stars in the sample. Fifty-one of our targets have a $v \sin i$ rms of zero m \pers, but for the purposes of this plot have been set at 50 m \pers ~for illustration purposes. We note the lower sensitivity limit of half each instrument's resolution for TRES (pale blue) and CHIRON (teal) as dashed lines. As expected, the rms increases for the fainter targets, especially those with detectable rotational broadening.  \label{fig:vsini_uncertainties}}
\end{figure}

Because we have four rotational broadening measurements for most of our stars, we can investigate the uncertainties on our measurements via the standard deviation in each $v \sin i$ value. For the minimally broadened subsample of effectively single stars, the root-mean-square (rms) of our $v \sin i$ measurements ranged from $0$ to $4.7$ km \pers, with a median rms of 0.72 km \pers. The targets with detectable rotational broadening exhibited $v \sin i$ rms values of $0.1$ to $10.4$ km \pers, with a median value of 0.42 km \pers. We show in Figure \ref{fig:vsini_uncertainties} the $v \sin i$ rms versus R magnitude for the effectively single targets, with targets with detectable rotational broadening indicated. The $v \sin i$ rms notably increases for the fainter targets in our sample.

Roughly one-third of our effectively single sample ($29\pm3$\%; 95 out of 328) have $v \sin i$ greater than half the spectral resolution of each instrument. We place upper limits of 1.9 km \pers ~(for stars observed with CHIRON) and 3.4 km \pers ~(for stars observed with TRES) on the remaining targets.  \citet{Browning(2010)} reported that roughly 30\% of the six stars in their sample with spectral types later than M4 V had $v \sin i$ larger than their detection limit of 2.5 km \pers. Our result is slightly larger than the 23\% of 328 nearby M dwarfs that exhibited significant rotational broadening (i.e., $v \sin i > 2.0$ km \pers) reported in \citet{Reiners(2018)}. The slight difference is not surprising, as that study included spectral types across the entire M-dwarf sequence, many of which are early M dwarfs that exhibit minimal rotational broadening.

While broadened spectral lines can indicate the rotational broadening of the star in the case of single stars, a composite spectrum due to an unresolved binary can also manifest as broadening. This is the case for 48 close multiple systems, where we measured $v \sin i$ larger than our instrumental detection limits. We identify these measurements with an asterisk in Table \ref{tab:sample_data} as a flag to use these values with caution, as we cannot distinguish between true rotational broadening and broadening due to overlapping spectral lines from a blended spectrum. For the remaining 34 close multiple systems, we measured $v \sin i$ values less than our detection limits. As with the effectively single stars in our sample, we adopt half the spectral resolution of the instrument used as the upper limit on the $v \sin i$ for these systems. These are all noted in Table \ref{tab:sample_data}.

\subsubsection{Radial Velocities}
\label{subsubsec:rv_msmts}

To calculate RVs, we use standard cross-correlation techniques based on the those described in \citet{Kurtz(1998)}. We have reported RVs in previous work \citep{Winters(2018),Winters(2020),Medina(2022b),Pass(2023a), Pass(2023b)}; however, as described in \citet{Pass(2023a)}, we have updated and improved our analysis to include additional echelle apertures and to utilize co-added shifted and stacked templates. We summarize these changes below.

Instead of using only one echelle aperture, as we have done in previous work \citep{Winters(2018),Winters(2020)}, we include six apertures in our radial velocity measurements. The six apertures (TRES: 36,38,39,41,43,45; CHIRON: 36,37,39,40,44,51) permit the calculation of per-spectrum RV uncertainties and were hand selected to be reasonably free of telluric lines in the red wavelength range of roughly $6400-8000$ \AA, which has high information content for low-mass M dwarfs. We exclude the aperture containing $H\alpha$ for our RV measurements.  

As noted above in \S \ref{subsubsec:vsini_msmnts}, we estimate an initial RV for each star using a high-SNR spectrum of Barnard's Star, as described in \citet{Winters(2020)}. Barnard's Star is a slowly rotating \citep[130.4 days,][]{Benedict(1998)} M4.0 dwarf \citep{Kirkpatrick(1991)} for which we adopt a Barycentric radial velocity of $-110.3\pm0.5\ {\rm km\ s^{-1} }$, derived from presently unpublished CfA Digital Speedometer \citep{Latham(2002)} measurements taken over 17 years. We see negligible rotational broadening in our Barnard's Star template, which is in agreement with an expected $v \sin i$ of 0.07 km \pers ~from its long photometric rotation period. This is also consistent with the $v \sin i$ upper limit of $2 ~{\rm km\ s^{-1} }$ reported by \citet{Reiners(2018)}. We then shift and stack the spectra of single, inactive stars in our sample to create templates based on the information content $Q$ in echelle aperture 45: six templates for our TRES data and three templates for our CHIRON data. The smaller number of templates created from the CHIRON data reflects the smaller number of stars observed with CHIRON (148 targets, compared to 262 targets observed with TRES). For each star, we choose as the best template the one that maximizes the cross-correlation coefficient. We refer the reader to \citet{Pass(2023a)} for more detail. 


We followed the same methodology for our multiple systems as for our effectively single systems when calculating their RVs. Thus, unresolved multiples without measured orbits will have a composite RV.

Two stars were too faint and/or rapidly rotating for reliable RV measurements: 2MA~J0943-3833 and LHS~2471, which have $V$ magnitudes of 17.10 and 18.10 mag. These two targets will need larger telescopes for RV measurements. The binary LP~320-416AB, with angular separation 4.5\arcsec, was not always cleanly resolved enough to calculate individual RVs for each component, so we report a composite RV for the system. Thus, we do not report RVs or $v \sin i$ measurements for 2MA~0943-3833, LHS~2471, or LP~320-416B, so the number of stars for which we report $v \sin i$s and RVs is 410 out of the 413 stars in our sample.

\begin{figure}
\includegraphics[scale=.40,angle=0]{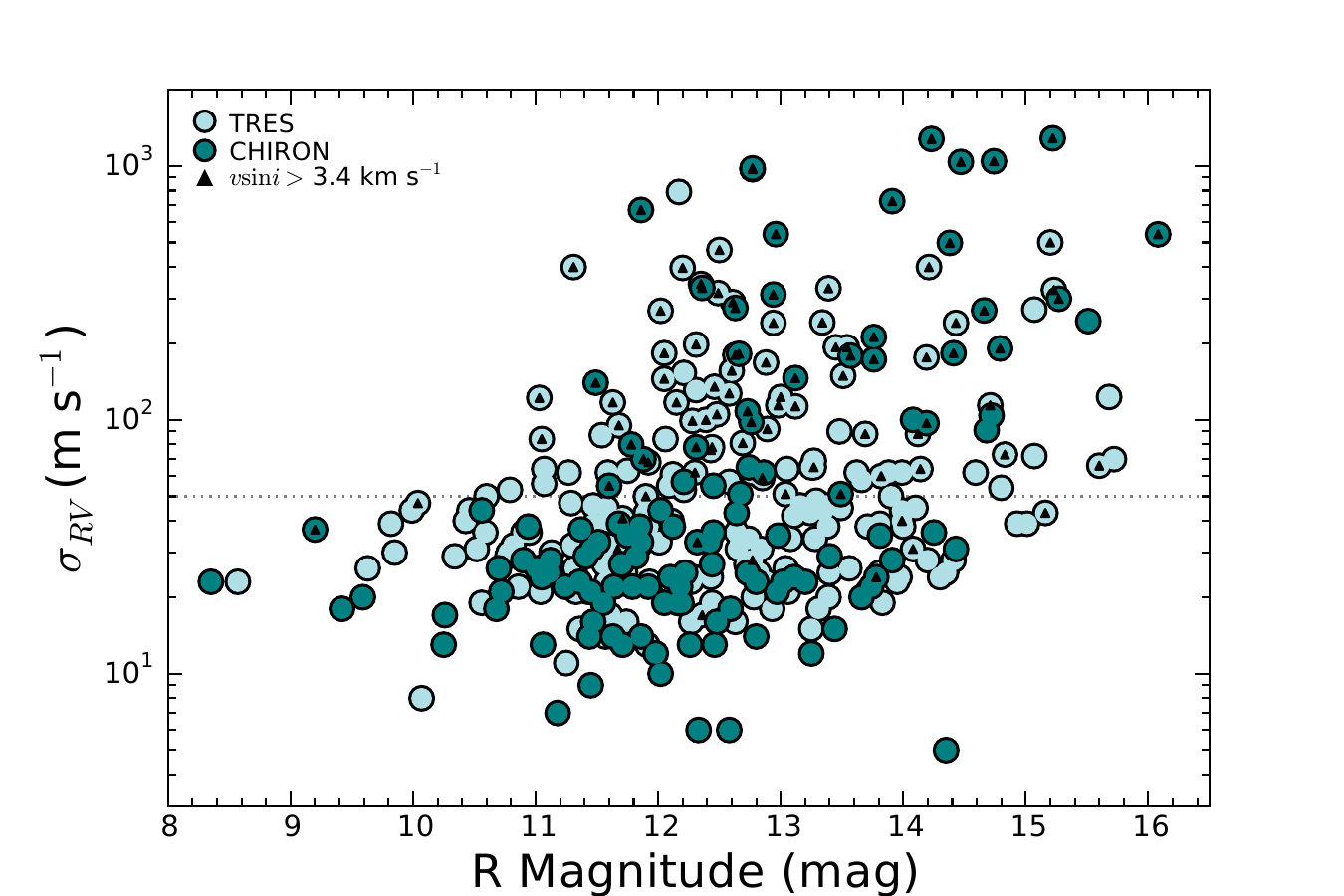}
\caption{RV rms as a function of $R-$band magnitude for the 328 presumed single stars in the sample. Our RV uncertainties are generally less than 1 km \pers ~for all of our stars and less than 50 m s$^{-1}$ (horizontal dotted line) for most of the stars with rotational broadening measurements below our detection limit ($v \sin i <$ 3.4 km \pers) The two minimally broadened outliers with large RV rms are GJ~512B and GJ~852A, which are  candidate multiples, as described in the text in \S \ref{subsec:new_mults}.  \label{fig:rv_uncertainties}}
\end{figure}

\subsubsection{Radial Velocity Uncertainties}
\label{subsubsec:rv_err}

As described by \citet{Pass(2023a)}, we calculate RV uncertainties using two different methods. We calculate theoretical uncertainties using the methods in \citet{Bouchy(2001)}, and we also calculate uncertainties directly from the cross-correlation coefficient using the methods in \citet{Zucker(2003)}. We conservatively adopt the larger uncertainty of the two methods. Following \citet{Tronsgaard(2019)}, we also add in quadrature an uncertainty term associated with our calculation of the barycentric correction at the geometric midpoint of the observation, although this contribution is generally negligible.

The uncertainties on our RVs range from $0.003$ to $0.127$ km \pers ~for the minimally broadened targets in our effectively single subsample, with a median value of 0.013 km \pers. For the subsample with detectable broadening, our RV uncertainties range from $0.007$ to $1.029$ km \pers, with a median value of 0.06 km \pers. 

The individual relative RVs for the first four observations of our effectively single targets in  the core spectroscopic sample were presented in \citet{Pass(2023a)} and \citet{Pass(2023b)}. Except for the five systems for which we are reporting spectroscopic orbits in a future paper (WT~766BC, GJ~164A, LP~69-457AB, GJ~376BC, and SCR~J0533-4257AB), we present our final list of individual RVs that include all epochs of data for our sample stars in Table \ref{tab:rv-data}.  We discuss the calculation of gamma velocities for the multiples in our sample in \S \ref{subsec:gamma_mult}. 

From our TRES and CHIRON spectra, we calculate the weighted mean RV and weighted mean internal uncertainty from the six echelle apertures. These are listed in columns 25 and 26 in Table \ref{tab:sample_data}.

We show in Figure \ref{fig:rv_uncertainties} our intra-star weighted mean RV uncertainties as a function of $R-$band magnitude for the 328 presumed single stars in the sample. Our intra-star RV uncertainties are generally less than 30 m \pers. We highlight the targets with detectable rotational broadening, and an expected increase in RV uncertainty is visible for these stars. We note that our RV uncertainties begin to creep above the 50 m \pers ~mark for the fainter ($R \gtrsim 14.5$ mag), slowly rotating targets.

As noted above, the CARMENES team and the Gaia DR3 reported RVs for some of our targets while our survey was in progress. We include the CARMENES results \citep{Lafarga(2020)} in Figure \ref{fig:us_lit_rv}  and perform an individual comparison to the Gaia DR3 results \citep{Katz(2023)} below. We acknowledge that more recent, noteworthy, high-resolution RV and $v \sin i$ measurements of mid-to-late M dwarfs within 15 pc have been published  \citep{Ishikawa(2022), Olander(2021)}, but an exhaustive comparison is beyond the scope of this study. We do not compare our measurements to results from any publications more recent than that of \citet{Lafarga(2020)}.

While Table \ref{tab:lit_spec} would indicate that 248 targets overlap with our sample, many of the targets observed by other surveys are the same well-known stars, so there are only 140 unique, effectively single stars for comparison, even when including more recent results from \citet{Jeffers(2018)} and \citet{Lafarga(2020)}. We illustrate in Figure \ref{fig:us_lit_rv} the difference between our RVs and those from the literature for the 124 effectively single stars in our sample with high-resolution (R $>$ 19,000) measurements. We specifically highlight comparisons to the four surveys with more than five targets in common with our sample. For the purpose of this comparison plot, we have adjusted the RVs from these four surveys to be on the same system as our RVs using a common star. This information is noted in Table \ref{tab:rv-comp_data} along with the range in RV differences and the median RV difference between our RVs and those of the four highlighted surveys. We chose only the most recent measurement for each star in our sample in this figure for comparison, which means that in many cases, results from \citet{Lafarga(2020)} take precedence. We are not able to perform a comparison to the RVs reported in \citet{Bonfils(2013)}, as those data are unavailable. The uncertainties on the RV differences shown in Figure \ref{fig:us_lit_rv} include our internal uncertainty, the 0.5 km \pers ~uncertainty on our Barnard's Star RV, and the literature RV uncertainty added in quadrature.

Illustrated is the excellent agreement between our RVs and those from \citet{Shkolnik(2012)}, \citet{Jeffers(2018)}, and \citet{Lafarga(2020)}, and moderate agreement with some scatter from those from \citet{Gizis(2002)}. The RVs from \citet{Gizis(2002)} were measured from the lowest resolution spectra in our comparison and thus it is not surprising that their RVs have large errors. Aside from the outliers in Figure \ref{fig:us_lit_rv}, it may appear that our uncertainties are overestimated. But this is only true if one is operating under the assumption that each individual point is independent and drawn from a Gaussian distribution. However, there is in some sense only four independent measurements here. Those are typically dominated by the systematic error in the scales of the instruments in each comparison publication (except in cases where observational errors are large and hence dwarf the 0.5 km \pers ~systematic error). In addition, many of the systematic effects that affect the RV zero point are stellar in origin (e.g., convective blue-shift, gravitational red-shift, stellar activity, magnetic cycles); thus, a simple comparison of the differences between literature values does not provide a full picture of the systematic error. It has been noted that absolute RV precision better than roughly $0.5$ km \pers ~is not possible unless a detailed physical model of the observed star is developed that takes into account these stellar effects \citep{Lindegren(2003), Lovis(2010), Lindegren(2021c)}.

\begin{figure}
\hspace{-0.25cm}
\includegraphics[scale=.40,angle=180]{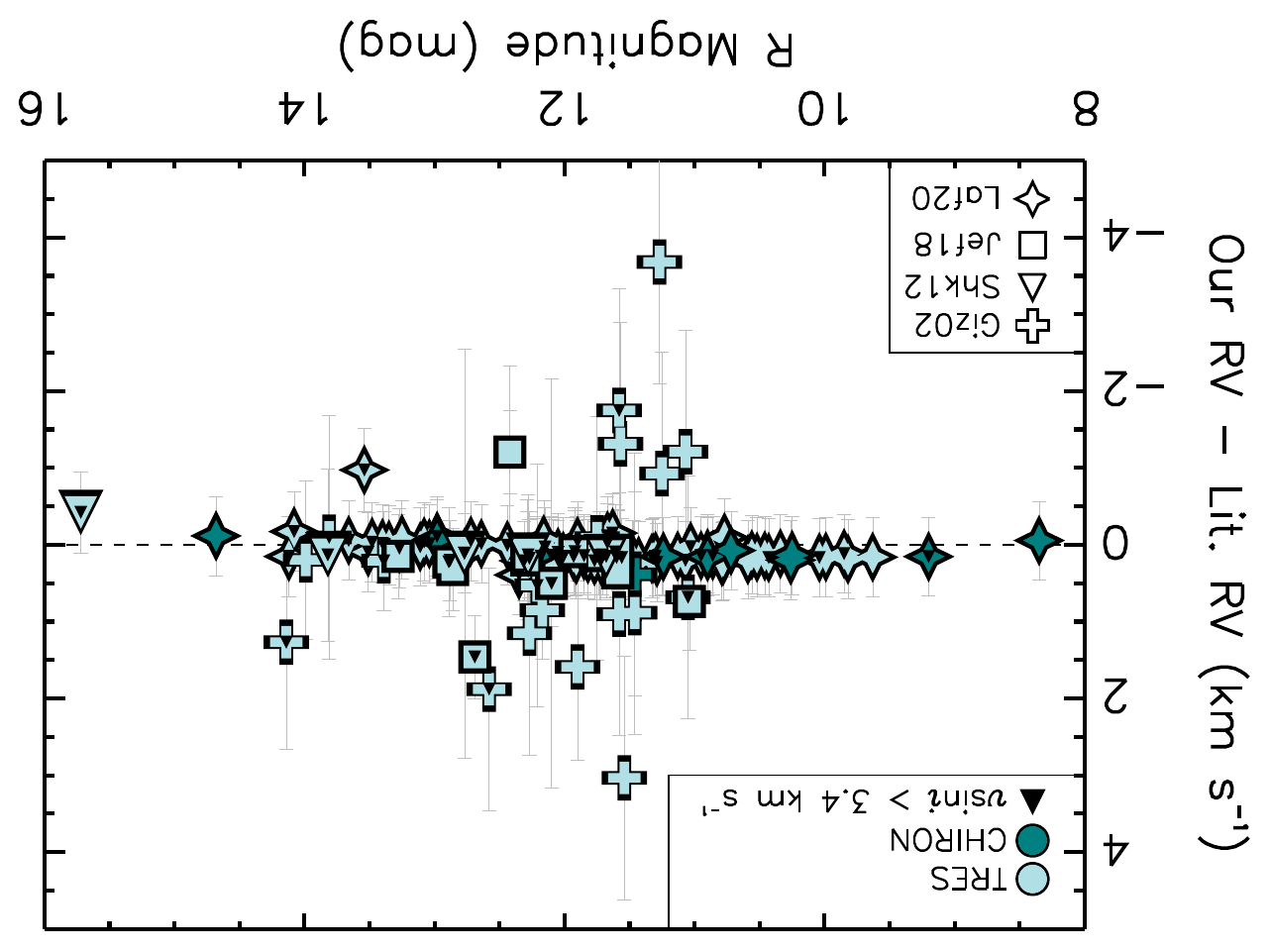}
\caption{RV difference as a function of $R-$band
magnitude for the 124 effectively single stars in our sample with high-resolution (R $>$ 19,000) measurements in the literature. We show a comparison to only those publications with more than five common targets \citep{Gizis(2002), Shkolnik(2012), Jeffers(2018), Lafarga(2020)}. TRES measurements are shown in pale blue; CHIRON measurements shown in teal. We also note the targets with $v \sin i >$ 3.4 km \pers. As noted in the text and in Table \ref{tab:rv-comp_data}, all RVs have been adjusted to be on the same system. The RV uncertainties shown illustrate our uncertainties, including the 0.5 km \pers ~from our Barnard's Star RV, and those from the literature added in quadrature.   \label{fig:us_lit_rv}}
\end{figure}

\citet{Katz(2023)} presented RV measurements from Gaia EDR3 \citep{GaiaDR3(2023)} spectra, of which 274 are for effectively single systems in our sample. We have adjusted the DR3 RVs to our Barnard's Star RV zero point and note that this RV difference of $+0.168$ km \pers ~is close to the correction of $+0.139$ km \pers ~needed to go from the native Center for Astrophysics $|$ Harvard \& Smithsonian Digital Speedometer velocity scale \citep{Latham(2002)} to the IAU absolute velocity scale. When taking into account the $0.5$ km \pers ~uncertainty on our Barnard's Star template, we can consider our RVs to be on the IAU absolute velocity scale. Most targets have RV differences within $\pm 2$ km \pers ~of our RVs or or uncertainties less than $2$ km \pers. The ones that are larger are all faint targets ($G >12.2$ mag ) and/or have detectably broadened spectra. We anticipate improved RV results from future Gaia data releases.

\begin{deluxetable}{lcccc}
\tabletypesize{\scriptsize}
\tablecaption{Individual Radial Velocities for 537 Nearby M Dwarfs \label{tab:rv-data}}
\tablecolumns{5}

\tablehead{
\colhead{Star Name} &
\colhead{BJD\tablenotemark{a}} & 
\colhead{RV\tablenotemark{b} } &
\colhead{$\sigma_{RV}$\tablenotemark{c} } &
\colhead{Instr.\tablenotemark{d}} \\
\colhead{} &
\colhead{(days)} & 
\colhead{(${\rm km\ s^{-1}}$)} &
\colhead{(${\rm km\ s^{-1}}$)} &
\colhead{} 
}
\startdata
GJ1001A           &2458329.8474 & 32.357   & 0.020 & C \\
GJ1001A           &2458707.8194 & 32.335   & 0.033 & C \\
GJ1001A           &2458723.7259 & 32.311   & 0.022 & C \\
GJ1001A           &2459185.6148 & 32.313   & 0.020 & C \\
GJ1002            &2457933.9686 & -39.868  & 0.029 & T \\
GJ1002            &2458106.6685 & -39.894  & 0.098 & T \\
GJ1002            &2458109.5924 & -39.878  & 0.031 & T \\
GJ1002            &2458689.9704 & -39.910  & 0.029 & T \\
G217-032AB        &2457678.8544 & 2.825    & 0.035 & T \\ 
G217-032AB        &2457932.9765 & 2.883    & 0.058 & T \\
\hline
\enddata
\tablenotetext{a}{Barycentric Julian Date of mid-exposure, in the TDB
  time-system.}
\tablenotetext{b}{Barycentric radial velocity.} 
\tablenotetext{c}{The internal model-dependent uncertainties on each listed velocity. These uncertainties do not include the 0.5 km \pers ~uncertainty associated with the RV of our Barnard's Star template.}
\tablenotetext{d}{Instrument. `C': CHIRON; `T': TRES. }
\tablecomments{The velocities for the first few systems in our sample are
  shown to illustrate the form and content of this table. The full
  electronic table is available in the online version of the paper.}
\end{deluxetable}

\begin{deluxetable*}{llccccr}
\tabletypesize{\scriptsize}
\tablecaption{Data \& Results for Radial Velocity Comparison \label{tab:rv-comp_data}}
\tablecolumns{7}
\tablehead{
\colhead{Publication} &
\colhead{Comparison Star} & 
\colhead{Our RV } &
\colhead{Literature RV } &
\colhead{RV Diff. Range} &
\colhead{Med. RV Diff. } &
\colhead{\# Stars} \\
\colhead{} &
\colhead{} & 
\colhead{(${\rm km\ s^{-1}}$)} &
\colhead{(${\rm km\ s^{-1}}$)} &
\colhead{(${\rm km\ s^{-1}}$)} &
\colhead{(${\rm km\ s^{-1}}$)} &
\colhead{} 
}
\startdata
\citet{Gizis(2002)}   &Barnard's Star & $-110.3\pm0.5$   & $-111.1\pm0.4$  & $6.7$ & $0.53$     & 20 \\
\citet{Shkolnik(2012)}&LP~71-82       & $-1.26\pm0.5$   & $-1.2\pm0.2$    & $0.6$  & $0.13$     & 5 \\
\citet{Jeffers(2018)} &Barnard's Star & $-110.3\pm0.5$   & $-110.12\pm0.14$  & $2.7$  & $0.20$ & 13 \\
\citet{Lafarga(2020)} &Barnard's Star & $-110.3\pm0.5$   & $-111.156\pm0.110$  & $1.4$  & $0.14$  & 86 \\
\citet{Katz(2023)}    &Barnard's Star & $-110.3\pm0.5$   & $-110.468\pm0.131$  & $10.3$  & $0.06$  & 274 \\
\hline
\enddata
\end{deluxetable*}

Our high-resolution spectroscopic data have significantly increased the coverage of the southern targets in our sample. Of the 153 effectively single, southern targets, only 25\% (39) had previous $v \sin i$ measurements and only 23\% (35) had previous high-resolution RV measurements. When considering our full, all-sky sample, we have increased from 138 to 410 the number of high-resolution, multi-epoch RV measurements and increased from 63 to 410 the number of $v \sin i$ measurements for these nearby fully-convective stars. These represent a 66\% and a 84\% improvement in high-resolution RV and $v \sin i$ measurements, respectively,  compared to when our survey began. The addition of recent results \citep{Katz(2023), Lafarga(2020), Fouque(2018), Kesseli(2018), Jeffers(2018)} improves the coverage for stars in our sample, but without our data, five and 222 stars in this  sample would lack radial and rotational velocity measurements, respectively.

\section{Results}
\label{sec:results}

\subsection{New Detections of Multiple Systems}
\label{subsec:new_mults}

One of the goals of our survey was to search for binary systems orbital periods that are roughly half of the duration of our study, that is p$_{orb} <$ 3 yr. We were able to detect binaries via two different methods: 1) through the detection of multi-lined spectra, which indicates the presence of a bright  companion or 2) through velocity variations of the star in question, which indicates the presence of a faint stellar or brown dwarf companion. 

To identify bright companions, we searched for multiple spectral lines using least-squares deconvolution (LSD) plots \citep{Donati(1997)}, as described in \citet{Winters(2018)}, for every observation. To identify objects with velocity variations, we calculated the rms of all epochs of RVs available for an object and plotted those against the rotational broadening measurements of each target. As noted above, spectral lines can be artificially broadened in the case of multi-lined spectra. Thus, we scrutinized the results from each observation and compared all available observations of a target, often more than once. We investigated systems with RV rms larger than roughly 100 m \pers ~for the targets with $v \sin i$ below 3.4 km \pers. For objects with more broadened spectra, we looked for those with RV rms that were similar to that of the known binaries in our sample. We frequently obtained additional data for any promising targets, i.e., those with large RV rms.

We list here interesting systems in our sample, ordered by RA, some of which are double- or triple-lined and some of which exhibit RV trends indicative of orbital periods longer than our program can measure. These longer period single-lined binaries will be ideal for astrometric follow-up. The new systems listed below are in addition to our previously reported spectroscopic orbits for LHS~1610Ab, GJ~1029AB, LP~655-43ABC, LTT~115AcB, LHS~1817Ab, GJ~268AB, ~2MA~J0930+0227AB, LP~734-34AB, G~123-45Ab, G~258-17AB, and LTT~7077AB \citep{Winters(2018), Winters(2020)}. 

For the double-lined binaries, we estimate mass ratios and gamma velocities using the Wilson method \citep{Wilson(1941)}. We measured the flux ratios (flux$_B$/flux$_A$) that we observe in our spectra using an internal version of \texttt{todcor}, (modeled after \citealt{Zucker(1994)}) using Barnard Star as our template, using methods  previously presented \citep{Winters(2020)}. We list the re-normalised unit weight error (RUWE) value from Gaia DR3, where a value larger than 2.0 hints that the system is multiple, as noted in \citet{Vrijmoet(2020)}.

\subsubsection{Multi-lined Systems}

\textit{SCR~J0143-0602} (01:43 -06:02) is a new double-lined binary with a 115-day orbital period at 21.3 pc. We observed doubled lines in its spectrum and calculate a flux ratio of 0.97, a mass ratio of 1.04, and a gamma velocity of 14.29 km \pers ~from our observations. It's Gaia DR3 RUWE value is an insignificant 1.6 because its components are equal-luminosity and thus it likely shows a minimal shift in its photocenter in the Gaia data. We note that this system is also singly-eclipsing in TESS data. We will present the spectroscopic orbit for this system in an upcoming publication.

\textit{SCR~J0533-4257} (05:33 -42:57) is a known binary \citep{Riedel(2018)} with a published astrometric orbit with an orbital period of 0.672 yr \citep{Vrijmoet(2022)}. We observe doubled lines in its spectrum and calculate a flux ratio of 0.6, a mass ratio of $0.89\pm0.02$, and a gamma velocity of $-3.10\pm0.08$ km \pers ~from our observations. The Gaia DR3 RUWE value is 11.8. We will present the spectroscopic orbit for this system in an upcoming publication. 

\textit{GJ~376B} (10:00 +31:55) was a suspected unresolved binary based on its overluminosity \citep{Gizis(2000),Raghavan(2010)}, and we confirm that it is double-lined with an orbital period of 945 d. We estimate a flux ratio of 0.7, a mass ratio of $1.03\pm0.04$ and a gamma velocity of $55.97\pm1.08$ km \pers ~from our spectra. This object has a RUWE value of 8.4 in the Gaia DR3. This new binary is part of a new hierarchical triple system at 14.5 pc with a solar-type primary component at a separation of 134\arcsec. We will present the spectroscopic orbit for this system in an upcoming publication.

\textit{LP~69-457} (16:40 +67:36) was reported as a binary candidate in \citet{Winters(2019a)} due to its overluminosity. We observe double lines in our spectra. We estimate a flux ratio of 0.5, a mass ratio of $0.72\pm0.03$  and a gamma velocity of $-17.93\pm0.40$ km \pers ~from our spectra. This system has a RUWE value of 9.8 in the Gaia DR3 and is a new binary system at 14.4 pc with an orbital period of 844 d. We will present the spectroscopic orbit for this system in an upcoming publication.

\textit{G~190-27} (23:29 +41:27) is a known subarcsecond binary \citep[0\farcs078;][]{Bowler(2015)}. We observe it to be double-lined in most of our observations, with a flux ratio of 0.8 a mass ratio of $0.91\pm0.01$ and a gamma velocity of $-14.45\pm0.15$ km \pers. It has a RUWE of 34.7 in Gaia DR3 and is part of a hierarchical triple where the more massive M dwarf primary G~190-28 is separated by roughly 18\arcsec. 

\textit{GJ~865} (22:38 -65:22) was a known subarcsecond binary \citep[0\farcs769;][]{Lindegren(1997)} at 12.8 pc that was later observed to be triple-lined [E. Jensen, private communication]. We also observe it to have triple lines and estimate flux ratios of 0.4 for both B/A and C/A. No astrometric information exists in the Gaia DR3. This is a possible new member of our core sample of mid-M dwarfs within 15 pc, as the presence of a third component reduces the previously estimated mass of the primary. 

\textit{GJ~792} (20:31 +38:33) is a known subarcsecond binary \citep[0\farcs118;][]{Janson(2014a)}. We see double lines in two epochs and estimate a flux ratio of 0.5, a mass ratio of $0.76\pm0.01$, and a gamma velocity of $-24.72\pm0.14$ km \pers ~from our data. The Gaia RUWE value is 34.3.

\subsubsection{Systems with Velocity Trends}
\label{subsubsec:rv_trends}

\begin{figure*}[ht!]
\includegraphics[scale=.35,angle=0]{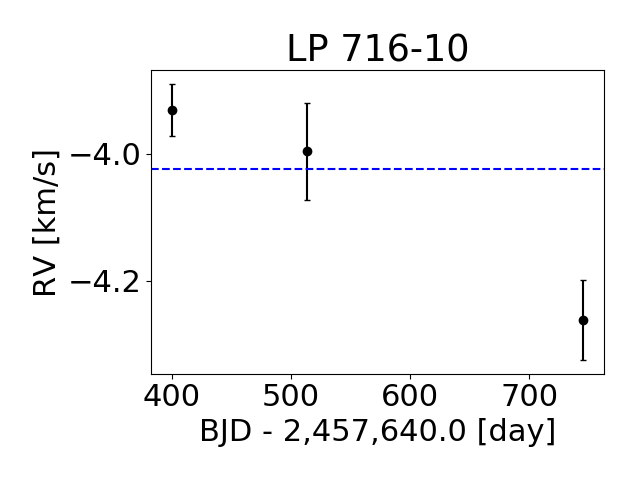}
\includegraphics[scale=.35,angle=0]{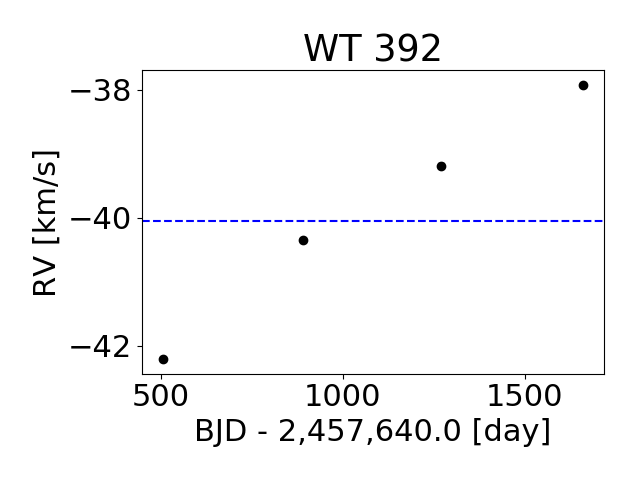}
\includegraphics[scale=.35,angle=0]{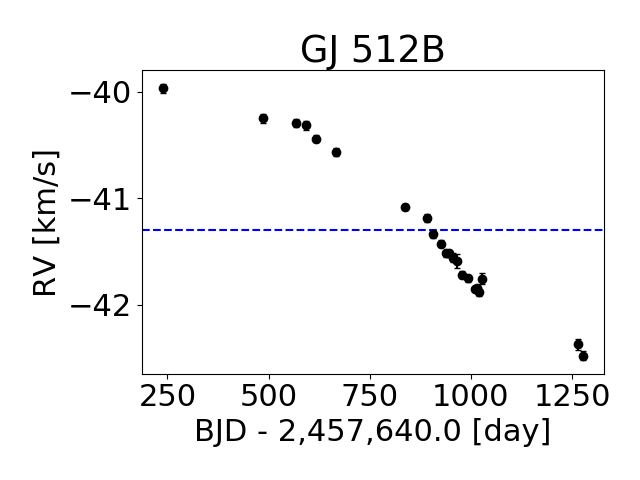}
\hspace{0.2cm}
\includegraphics[scale=.355,angle=0]{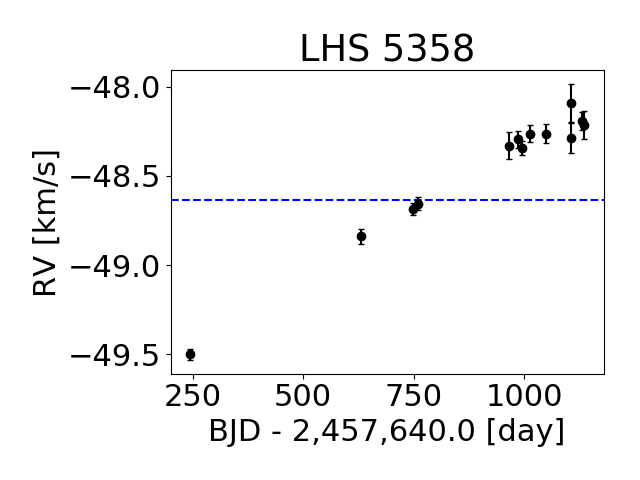}
\hspace{0.2cm}
\includegraphics[scale=.35,angle=0]{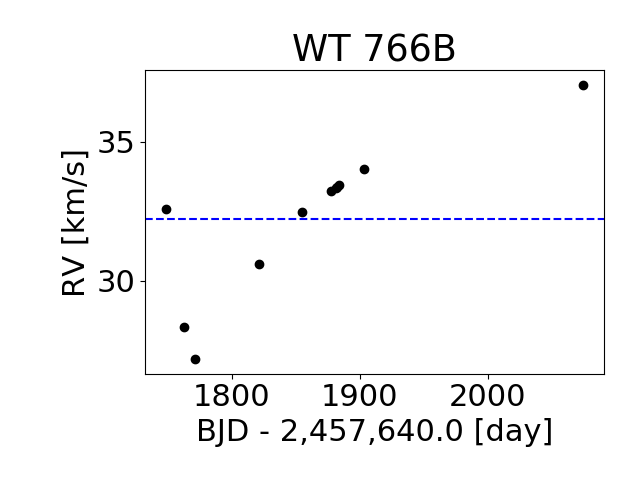}
\hspace{0.2cm}
\vspace{-0.35cm}
\includegraphics[scale=.35,angle=0]{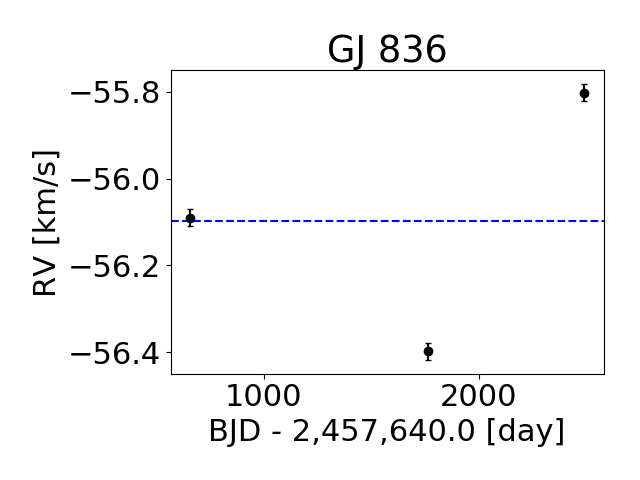}
\caption{The six systems in which we detect velocity trends indicative of a companion in a long-period orbit. The blue horizontal line indicates the weighted mean RV for each system.  \label{fig:rv_trends}}
\end{figure*}

Six systems show velocity trends that indicate the presence of an unseen companion. We show in Figure \ref{fig:rv_trends} the RVs for each system as a function of time and discuss each system below.

\textit{LP~716-10} (04:52 $-$10:58) shows a velocity trend of 354 m \pers ~over roughly one year, with an RV rms of 175 m \pers. The Gaia DR3 RUWE value is 16.8. This is a new binary detection at 16.3 pc. 

\textit{WT~392} (13:13 $-$41:30) was reported as a speckle detection in \citet{Vrijmoet(2022)} with an angular separation between the components of 0\farcs06. We measure an RV trend of 4.3 km \pers ~over roughly 3 yr and an RV rms of 1.8 km \pers. The Gaia DR3 RUWE value is 13.6. We will present the spectroscopic orbit for this system in an upcoming publication.

\textit{GJ~512B} (13:28 $-$02:21) exhibits velocity variations of 2.4 km \pers ~over nearly three years of observations, with an RV rms of 700 m \pers. This is not likely due to the early-M-type primary component separated by 8\arcsec ~unless we have serendipitously observed the A-B system at periastron. We estimate the semi-amplitude for the A-B orbit to be 1.8 km \pers, based on an estimated orbital period of 1500 yr and assuming inclination of 90\arcdeg ~and eccentricity of zero. The Gaia DR3 RUWE value is 7.9, which supports our suggestion of a new companion. This is a new candidate at 13.9 pc, and would mean that the GJ~512 system is a hierarchical triple should the candidate be confirmed.

\textit{LHS~5358} (20:43 $+$04:45) We detect a linear velocity trend of 1.4 km \pers ~over 2.4 years, with an RV rms of 380 m \pers. This system has a reported RUWE of 10.2 in Gaia DR3 and is a new binary detection at 15.4 pc.

\textit{WT 766Bc} (21:01  $-$49:07) has an astrometric orbit in the Gaia DR3 non-single star catalog \citep{GaiaArenou(2023)} with an orbital period of 597 d. We independently identified this target as a single-lined spectroscopic binary in an eccentric (e $=$ 0.6) 595-day period. The Gaia DR3 orbital elements have since been  combined with measurements from GRAVITY, a near-infrared interferometer at the VLT, to measure a companion's mass of 77 M$_{Jup}$ \citep{Winterhalder(2024)}. We will present the spectroscopic orbit for this system in an upcoming publication.

\textit{GJ~836} (21:39 $-$24:09) is reported as having an astrometric perturbation in \citet{Vrijmoet(2020)}; we observe a velocity variation of just over 600 m \pers ~with three observations over roughly three years. Its Gaia DR3 RUWE value is 1.5, and its distance is 19.5 pc.

For the seven newly detected multiples described above (LP~69-457AB, WT~766Bc, GJ~376BC, GJ~512Bc, SCR~J0143-0602AB, LHS~5358Ab, and GJ~836A), we note a separation placeholder of 1\arcsec ~in Table \ref{tab:sample_data}, which is surely an overestimate for these systems.

\subsubsection{Interesting Systems}

\textit{LTT~10491} (01:22 $+$22:09) is a known subarcsecond binary \citep{Cortes-Contreras(2014)} with a $\Delta I$ of 0.86 mag and a separation of 0\farcs271 between its components. This system has no Gaia parallax or RUWE parameter. We see a change in velocity on the order of 1 km \pers ~over the first three observations, which were taken over roughly three years, but subsequent observations have decreased the significance of this variation. We estimate an orbital period of roughly 9 years from its ground-based parallax of $87.60 \pm 8.50$ mas and deblended component masses of $0.188\pm0.026$ and $0.159\pm0.026$ \msun. It is possible that we serendipitously observed this system near periastron.

\begin{figure}
\hspace{-0.5cm}
\includegraphics[scale=.31,angle=0]{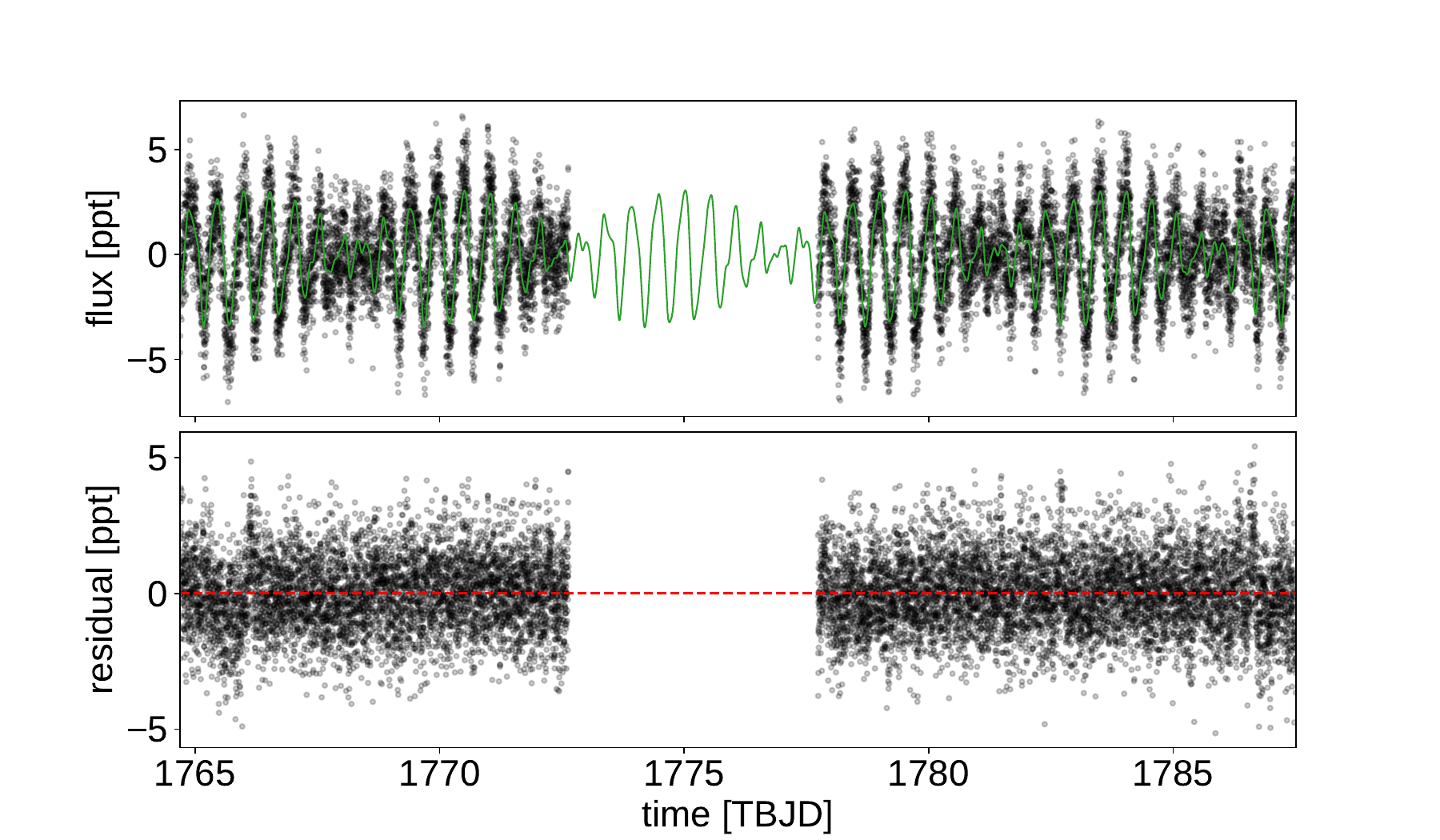}
\includegraphics[scale=.52,angle=0]{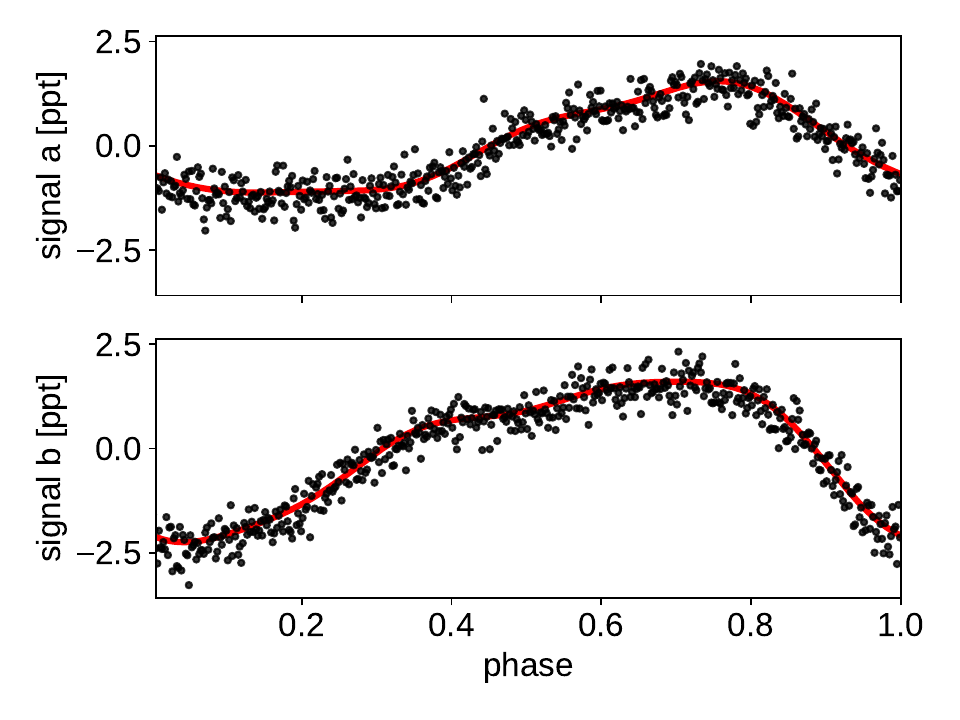}
\caption{Light curve and fit rotation periods from Sector 17 TESS data of LTT~10491. We observe the same two rotation periods in Sectors 42 and 57, but as the spot pattern evolves over time, we show only a single sector in this plot for clarity. Our best-fit model is shown in green in the upper panel and red in the lower two panels. This model includes sinusoids at periods of 0.56d and 0.50d, as well as sinusoids for the first three harmonics of each period. The upper panels include all data points; in the lower panels, we show 500 bins evenly spaced in phase, with the model of the second signal subtracted.\label{fig:ltt10491_prot}}
\end{figure}

Because we have measured the rotational broadening of this binary, we can explore the system's inclination if a photometric rotation period also is detectable. This target was observed by TESS at short-cadence (120 s) in sectors 17, 42, and 57 \citep[Program: G022076, PI: Winters;][]{LTT10491_TESS}. We used methodologies described in \citet{Pass(2023b)} to search for signals of rotational modulation, and we detect two rotation periods: 0.50 d and 0.56 d. We show the light curve and fits for the sector 17 data in Figure \ref{fig:ltt10491_prot}. 

Using the primary component's deblended mass of $0.229\pm0.015$ \msun, we calculate a stellar radius of $0.246\pm0.016$ \rsun ~using the empirical mass-radius relation for single stars in \citet{Boyajian(2012)}. Based on this radius and our measured photometric rotation periods, we expect a rotational velocity of $25$ km \pers ~if the primary is responsible for the 0.50-day or a rotational velocity of $22$ km \pers ~if it has a 0.56-day period. Our measured \vsini ~of $19.3$ km \pers ~would produce $\sin i > 0.77$ (assuming our \vsini ~measurement is not inflated by line blending; the fact that we do not see time variation in our \vsini~measurement supports this assumption), which indicates a rotation axis inclination $> 50$\arcdeg.

\textit{GJ~275.2} (07:30 $+$48:11) is a widely-separated (103\arcsec) companion to a pair of white dwarfs in a 20.5-yr orbit \citep{Harrington(1981)}. \citet{Harrington(1981)} reported this mid-M dwarf companion as an astrometric binary, and published a preliminary astrometric orbit with an orbital period of 0.9 years, eccentricity of 0.6, inclination of 94\arcdeg, and component masses of 0.17 \msun ~and 0.08 \msun, respectively. The orbital period is confirmed in \citet{Heintz(1994)} with a similar sized astrometric perturbation, but with a conflicting orbital directional motion. 

The Gaia RUWE of 1.4 does not present convincing evidence of an astrometric perturbation. This target has been observed with the technique of lucky imaging  \citep{Janson(2014a), Cortes-Contreras(2017)}, but neither group detected a companion above their respective detection limits of 100 mas and 150 mas. The expected angular separation from the preliminary orbit for this system at 11.29 pc is 0\farcs05. Thus, it is not surprising that a companion has not been detected via high-resolution imaging. 

\citet{Harrington(1981)} note that an RV variation on the order of 10 km \pers ~would be detected, but we see no velocity variations with 11 observations taken over six years. The inclination from the preliminary orbit indicates that the system is not in a face-on configuration. Higher cadence spectroscopic monitoring is needed to determine if this system is truly quadruple, rather than triple.

\textit{GJ~852} (22:17 $-$08:48) This is a known hierarchical triple system, where the A-BC separation is 7\farcs77 \citep{Skrutskie(2006)} and BC is a subarcsecond binary with a separation 0\farcs97 \citep{Bergfors(2010)}. The RUWE value for the B component is 1.8, while that of the A component is 10.1. Because we measure a similar RV rms for both the A and BC components: 0.74 km \pers ~and 0.86 km \pers, respectively, for our measurements over five years, we therefore suspect that the primary component is also a binary, but we do not see more than one set of lines in the spectrum. A background companion was detected via IR AO \citep{Vogt(2015)}. Additional spectroscopic monitoring of the A component is necessary to determine whether the primary component is itself an unresolved pair.

\subsection{Gamma Velocities}
\label{subsec:gamma_mult}

One of our goals is to calculate 3-d galactic space motions for our entire sample, for which we need the systemic or gamma velocity of the system. This is straightforward for the 285 presumed single systems, and we adopt as the gamma velocity our weighted mean RV. The adopted RV uncertainty listed in column 35 of Table \ref{tab:sample_data} is our weighted mean RV uncertainty added in quadrature with the 0.5 km \pers ~from our Barnard's Star template. These stars are noted with the code `s' in  the code column for the adopted gamma velocity in Table \ref{tab:sample_data}. 

The determination of gamma velocities for the multiple systems is challenging for the systems without measured orbits (i.e., mostly the longer period multiples). We generally have at least four observations for our targets over five (or fewer) years, but orbital periods for our multiples can range from hours to thousands of years. We describe below our method for determining the gamma velocity and related uncertainty for the multiples in our sample.

\begin{figure}
\includegraphics[scale=.42,angle=0]{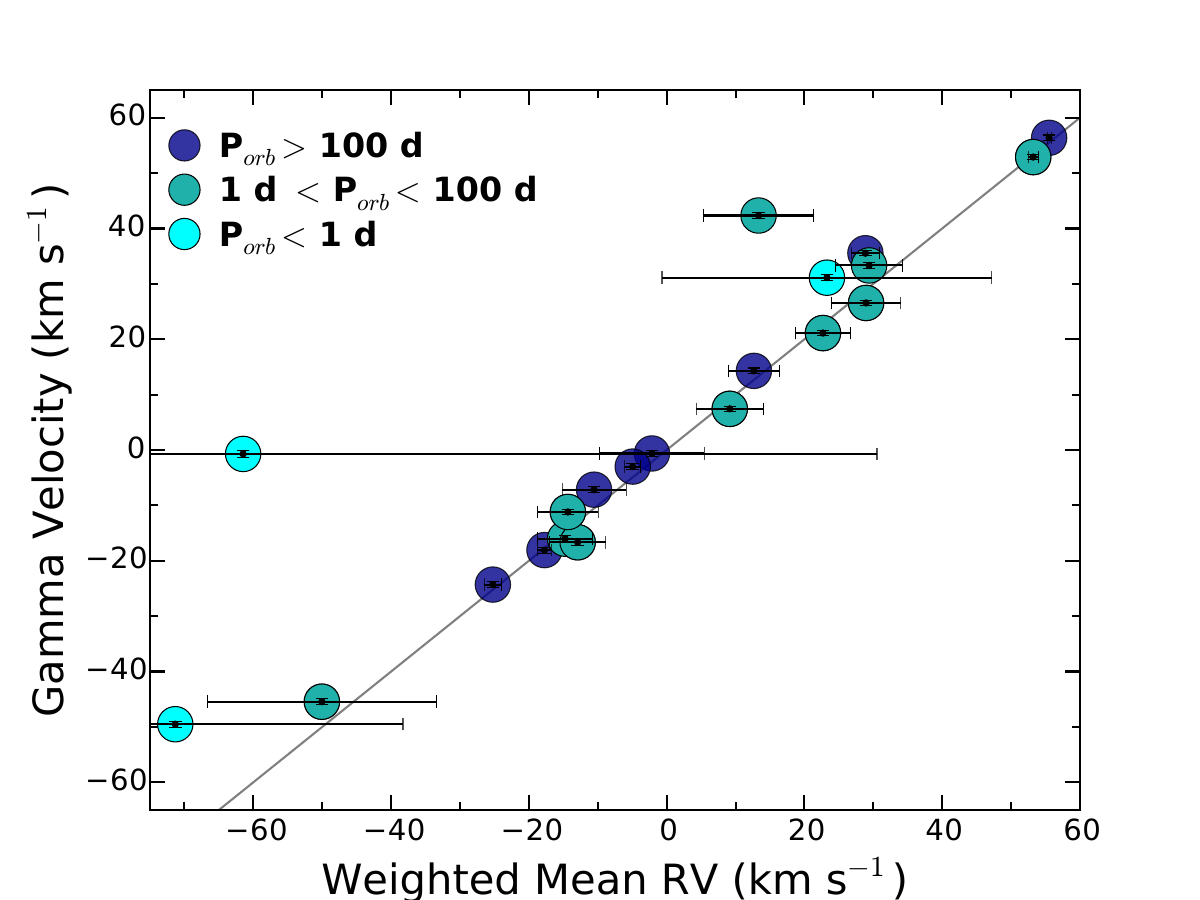}
\caption{Comparison of gamma velocity to weighted mean RV for spectroscopic multiple systems with published or finalized orbits from our program (\citealt{Winters(2018), Winters(2020)}, Winters, in prep). The orbital periods range from 0.3 d to 944 d. We use only the first four epochs of data to calculate the plotted weighted mean RVs. The uncertainties plotted are the standard deviation of the first four individual RV measurements added in quadrature with the 0.5 km \pers ~uncertainty on our Barnard's Star template RV. The gray line indicates 1:1 agreement. The three binaries with the shortest orbital periods (periods $<$ 1 day, in cyan) fall far from the 1:1 line, while most of the binaries with longer orbital periods (periods $>$ 1 day, teal and dark blue) lie very close to the 1:1 line. Thus, we conclude that while it is reasonable to adopt the weighted mean RV as the gamma velocity for the long-period multiples in our sample, we must -- and do -- adopt the gamma velocity from published orbits for the shortest period multiples, as we describe in the text. \label{fig:binary_err}}
\end{figure}

We want to understand how the true gamma velocity compares to our weighted mean RVs for our multiple systems, so we consider the 21 spectroscopic binaries for which we have published or finalized orbits from this program (\citealt{Winters(2018), Winters(2020)}; Winters, in prep). Their orbital periods vary from 0.3 d to 944 d. For these multiples, we compare the weighted mean RVs of the first four RV measurements for each system to our published or finalized gamma velocities (see Figure \ref{fig:binary_err}). Binaries with measured orbital periods longer than 100 days nearly all fall on the one-to-one line, save for WT~766BC. WT~766BC is unusual in that it is both eccentric ($e$ roughly 0.6) and our first four observations occurred just before and after periastron. Thus, it is no surprise that our weighted mean RVs for this system differs from its gamma velocity.

In short, Figure \ref{fig:binary_err} illustrates that the uncertainties on the RVs of the spectroscopic binaries in our sample with very short orbital periods will be underestimated if we consider only the weighted mean uncertainty. Thus, for very short period binaries, we must adopt the gamma velocity from the published orbital fit. We note that our mean RVs will not be correct for any unknown very short period binaries. Figure \ref{fig:binary_err} also illustrates that it is reasonable to adopt the weighted mean RV as the gamma velocity for the multiples in our sample with orbital periods p$_{orb} >$ 100 d.

We have measured the spectroscopic orbits for five systems which we will report in a future publication: WT~766BC, GJ~164A, LP~69-457AB, GJ~376BC, and SCR~J0533-4257AB. We include their gamma velocities and uncertainties from our orbital fits in the following description of how we have treated the systemic velocities for the multiple systems in our core 15 pc sample. 

Seventeen systems (16 binaries and one triple) have published spectroscopic orbits. For these, we used the published gamma velocity and its uncertainty when calculating the $UVW$ space motions. We include our gamma velocity and its uncertainty from our unpublished orbital fits for GJ~164A, LP~69-457AB, and SCR~J0533-4257AB in this group. These are noted with the code `o' in the code column for the adopted gamma velocity.

Four systems without full spectroscopic orbits are double- or triple-lined in our spectra. GJ~792AB is an SB2, so we use the Wilson method \citep{Wilson(1941)} to estimate the gamma velocity. The gamma velocity uncertainty of 0.52 km \pers ~includes the 0.5 km \pers ~uncertainty on our Barnards Star template. GJ~792 is noted with the code `w' in the code column for the adopted gamma velocity in Table \ref{tab:sample_data}. LTT~11586AcB is an ST2, while GJ~487ABC and LTT~12352ABC are ST3's, and published orbits are available for the inner pair of these three systems. For these three systems, we use our in-house version of \texttt{todcor} and \texttt{tricor} to calculate the individual velocities of each component at each observation. We then average all components' RVs over all epochs and use this averaged RV as the adopted gamma velocity. We adopt the rms of the measurements as the uncertainty on this velocity. These three systems are noted with the code `t' in the code column for the adopted gamma velocity in Table \ref{tab:sample_data}.

Fifty-two of our mid-M dwarf systems (51 binaries and one triple) are known, unresolved, multiple systems with composite spectra but without published spectroscopic orbits. These systems are expected to have orbital periods on the order of dozens to hundreds of years. Thus, we expect these systems to have small RV semi-amplitudes. Based on the longer period binaries (p$_{orb} > 100$ d) indicated in Figure \ref{fig:binary_err}, we do not expect that our weighted mean RVs will differ significantly from their gamma velocities. Thus, we adopt our weighted mean RVs as the gamma velocities, but we conservatively include the standard deviation of the RV instead of our internal error for the RV uncertainty, which we then add in quadrature with the 0.5 km \pers ~from our Barnard's Star template. These are noted with the code `c' in the code column for the adopted gamma velocity. 

This inflation of the RV uncertainty results in a significant RV uncertainty of between $1.19$ km \pers ~and $6.19$ km \pers ~for nine binary systems. The object with the largest RV rms, SCR~J2049, is a known short-period (1.77 yr) binary with a published astrometric orbit \citep{Henry(2018)}, but no spectroscopic orbit. All but one of these systems have primary masses $< 0.2$ M$_{\odot}$ and exhibit rotational broadening of more than 15 km \pers. Three of these systems are also fainter than $R = 14.6$ mag and thus have low SNR spectra.

An additional 53 systems contain one or more of our mid-M dwarf components as members of widely-separated multiple systems. These systems will thus have very long-period orbits (on the order of hundreds to hundreds of thousands of years). In nine cases we measured the RV of the primary as part of our survey. 

Five of these 53 multiple systems lack high-resolution RVs for their widely separated companion. This includes GJ~1001, a triple system with two widely-separated brown dwarf components, and four widely separated systems with white dwarf primaries -- GJ~754.1, GJ~630.1, WT~765 and GJ~283. In two cases (GJ~630.1 and WT~765), the mid-M dwarf is part of a closely separated pair for which there is a spectroscopic orbit and it is the gamma velocity for this pair that we adopt as the systemic velocity for the triple. We include the gamma velocity and uncertainty from our unpublished orbital fit for WT~766BC in this group. For the other three, we adopt our measured mid-M dwarf weighted mean RV and its uncertainty as the gamma velocity and uncertainty and note all five with the code `a' in  the code column for the adopted gamma velocity in Table \ref{tab:sample_data}. 

For seven of the 53 systems, one pair or triple of a hierarchical system has a spectroscopic orbit with a gamma velocity. We include the gamma velocity and uncertainty from our unpublished orbital fit for GJ~376BC in this group. We take the weighted mean of that value with the individual RV value of the remaining stellar component(s) as the systemic velocity. These are noted with the code `n' in  the code column for the adopted gamma velocity in Table \ref{tab:sample_data}.

We use published RVs for the remaining 32 primary components of the widely separated multiple systems. The majority (23) of these published RV measurements are from the Gaia collaboration \citep{Katz(2023),Soubiran(2018)}. As we note in the comparison of our RVs to the Gaia DR3 RVs in \S \ref{subsubsec:rv_msmts}, we adjusted the Gaia RVs to our Barnard's Star zero point. The median difference between the weighted mean RV of the mid-M dwarf companion and their more massive primaries with Gaia RVs is $0.56$ km \pers. For these remaining 32 multiple systems where each component has an individual velocity, we took the weighted mean of the two components' velocities and velocity uncertainties as the gamma velocity and its uncertainty. These are noted with the code `m' in the code column for the adopted gamma velocity in Table \ref{tab:sample_data}.

We list in Table \ref{tab:wide_mults} the 34 systems with RVs of the primaries from the literature and list the weighted mean RVs for the mid-M dwarf components but the value in the RV uncertainty column reflects our intra-star RV rms added in quadrature with the 0.5 km \pers ~uncertainty on our Barnards Star template. 

\begin{deluxetable*}{lrcclrccr}
\tabletypesize{\scriptsize}
\tablecaption{RVs of Wide Multiple Components with Non-Mid-M Dwarf Primaries \label{tab:wide_mults}}
\tablecolumns{9}
\tablewidth{0pt}
\tablehead{\colhead{Mid-M Dwarf Name}        &
    \colhead{RV}          &
    \colhead{$\sigma_{RV}$}           &
    \colhead{Ref}          &
    \colhead{Primary Star Name}          &
    \colhead{RV}          &
    \colhead{$\sigma_{RV}$}           &
    \colhead{Ref} &
    \colhead{$\Delta$ RV}
}
\startdata
GJ~15B         &    11.14 & 0.50  &   1  &    GJ~15A       &   11.37 & 0.12 &    3   & -0.23 \\
GJ~51(B)       &    -5.88 & 0.61  &   1  &    GJ~49(A)     &   -6.28 & 0.13 &    3   &  0.40 \\
GJ~61B         &   -28.01 & 0.50  &   1  &    GJ~61A       &  -28.75 & 0.13 &    3   &  0.74 \\
LHS~1376(B)    &    18.77 & 0.50  &   1  &    LHS~1377(A)  &   18.36 & 0.21 &    3   &  0.41 \\
GJ~105B        &    26.36 & 0.50  &   1  &    GJ~105A      &   25.21 & 0.14 &    3   &  1.15 \\
LP~993-116(BC) &    28.42 & 0.50  &   1  &    LP~993-115(A)&   27.90 & 0.24 &    3   &  0.52 \\
GJ~166C        &   -43.18 & 0.50  &   1  &    GJ~166A      &  -42.47 & 0.12 &    3   & -0.71 \\
GJ~166C        &   -43.18 & 0.50  &   1  &    GJ~166B      & \nodata  &\nodata&\nodata &\nodata \\
LSPM~J0536(B)    &    21.95 & 0.50  &   1  &    GJ~208(A)    &   21.36 & 0.13 &    3   &  0.59 \\
G~192-12(B)    &     0.27 & 0.51  &   1  &    G~192-11(A)  &   -0.29 & 0.15 &    3   &  0.56 \\
GJ~275.2AD     &   -56.28 & 0.50  &   1  &    GJ~275.2BC   &  -56.49 & 0.05 &    8   &  0.20 \\
GJ~283B        &   -28.60 & 0.50  &   1  &    GJ~283A      & \nodata  &\nodata&\nodata &\nodata \\
GJ~324B        &    27.56 & 0.50  &   1  &    GJ~324A      &   27.19 & 0.12 &    3   &  0.37 \\
GJ~334B        &    37.79 & 0.71  &   1  &    GJ~334A      &   36.50 & 0.13 &    3   &  1.29 \\
GJ~376BC       &    56.37 & 0.51  &   9  &    GJ~376A      &   55.99 & 0.12 &    3   &  0.38 \\
GJ~412B        &    69.58 & 0.50  &   1  &    GJ~412A      &   68.41 & 0.12 &    3   &  1.17 \\
GJ~442B        &    17.21 & 0.50  &   1  &    GJ~442A      &   16.94 & 0.12 &    3   &  0.27 \\
L~758-107(B)   &    -9.13 & 0.50  &   1  &    L~758-108(A) &   -9.31 & 0.16 &    3   &  0.18 \\
GJ~512B        &   -41.30 & 0.87  &   1  &    GJ~512A      &  -39.40 & 0.50 &    1   & -1.89 \\
GJ~1179B       &   -12.41 & 0.50  &   1  &    GJ~1179A     &  -13.23 & 1.87 &    3   &  0.82 \\
PROXIMA~CEN    &   -21.50 & 0.50  &   1  &    $\alpha$~CEN &  -22.33 & 0.01 &    4   &  0.83 \\
GJ~611B        &   -58.87 & 0.50  &   1  &    GJ~611A      &  -59.44 & 0.12 &    3   &  0.57 \\
GJ~618B        &    29.95 & 0.50  &   1  &    GJ~618A      &   27.96 & 0.16 &    3   &  1.99 \\
GJ~630.1AB     &  -118.6  & 0.8   &   5  &    GJ~630.1C    & \nodata  &\nodata&\nodata &\nodata \\
GJ~643(D)      &    15.96 & 0.50  &   1  &    GJ~644ABE    &   14.95 & 0.01 &    7   &  1.01 \\
GJ~643(D)      &    15.96 & 0.50  &   1  &    GJ~644C      &   14.49 & 0.50 &    1   &  1.46 \\
GJ~669B        &   -34.51 & 0.51  &   1  &    GJ~669A      &  -35.03 & 0.51 &    1   &  0.53 \\
LHS~461(B)     &    -0.49 & 0.50  &   1  &    LHS~462(A)   &    0.45 & 0.25 &    3   & -0.94 \\
GJ~725B        &     1.13 & 0.50  &   1  &    GJ~725A      &   -0.63 & 0.50 &    1   &  1.76 \\
GJ~754.1B      &    11.80 & 0.50  &   1  &    GJ~754.1A    & \nodata  &\nodata&\nodata &\nodata \\
GJ~770C        &    -5.26 & 0.50  &   1  &    GJ~770AB     &   -5.27 & 0.01 &    2   &  0.01 \\
GJ~774B        &    28.17 & 0.50  &   1  &    GJ~774A      &   28.05 & 0.17 &    3   &  0.12 \\
WT~766(BC)     &    35.53 & 0.01  &   9  &    WT~765(A)    & \nodata  &\nodata&\nodata &\nodata \\
L~645-73(B)    &    26.78 & 0.50  &   1  &    L~645-74(A)  &   26.02 & 0.15 &    3   &  0.76 \\
LP~876-10(C)   &     6.43 & 0.64  &   1  &    FOMALHAUT(AB)&    6.50 & 0.50 &    6   & -0.07 \\
GJ~896B        &     3.68 & 0.51  &   1  &    GJ~896A      &    0.57 & 0.50 &    1   &  3.12 \\
\enddata
\tablerefs{
(1) this work; 
(2) \citet{Fekel(2017)};
(3) \citet{Katz(2023)};
(4) \citet{Kervella(2017)};
(5) \citet{Lacy(1977)};
(6) \citet{Mamajek(2013)};
(7) \citet{Segransan(2000)};
(8) \citet{Soubiran(2018)};
(9) Winters et al., in prep.
}
\tablecomments{Letters in parenthesis after the star name indicate the component for multiple system components with differing identifiers. }
\end{deluxetable*}

We present in Table \ref{tab:sample_data} the weighted mean RVs and their uncertainties of all of our measurements for each system in our sample. In the cases where the system is a known binary for which an orbit has been published, we note the gamma velocity in the column for literature measurements. We list in Table \ref{tab:sample_data} our adopted value and uncertainty for the gamma velocities, in addition to the notes described above. We note that in the cases where the best RV is that from our program, the \textit{adopted} RV uncertainty for each target noted in column 35 of Table \ref{tab:sample_data} includes the $0.5$ km \pers ~uncertainty on our Barnard's Star template, added in quadrature to our internal weighted mean uncertainties from column 26 of Table \ref{tab:sample_data}.

\subsection{UVW Space Motions}
\label{subsec:uvw}

When combined with proper motions and parallaxes, our precise RVs permit the calculation of accurate three-dimensional space motions. These allow estimates of membership to the thin disk, thick disk, and halo populations, which are a proxy for age. Old stars are inactive and have large space velocities due to passes through the spirals of the Milky Way, while young stars are generally active and have small space velocities \citep{Binney(1998)}.

We use each object's proper motion, parallax, and systemic velocity, along with each parameter's associated uncertainty, to calculate $UVW$ space motions and their uncertainties for our sample. We follow the methodology of \citet{Johnson(1987)}, updated to the  International Celestial Reference System (ICRS) frame of reference using the description in \citet{Perryman(1997)}. We use a right-hand coordinate system, where $U$ is positive in the direction of the galactic center, $V$ is positive in the direction of galactic rotation, and $W$ is positive in the direction of the northern galactic pole. Our median uncertainties in all three $UVW$ velocity components are 0.24 km \pers ~and are dominated by the uncertainty on our Barnard's Star template. When considering the median uncertainties in $UVW$ for the multiples in our sample, these are slightly larger: (0.29, 0.26, 0.23) km \pers. We then correct these motions to the local standard of rest (LSR) using the solar values of $(UVW)_{\odot}=$ (11.10, 12.24, 7.25) km \pers ~from \citet{Schonrich(2010)}. We list in columns $37-42$ in Table \ref{tab:sample_data} the uncorrected $UVW$ velocities and their uncertainties.

\begin{figure*}
\includegraphics[scale=.3,angle=0]{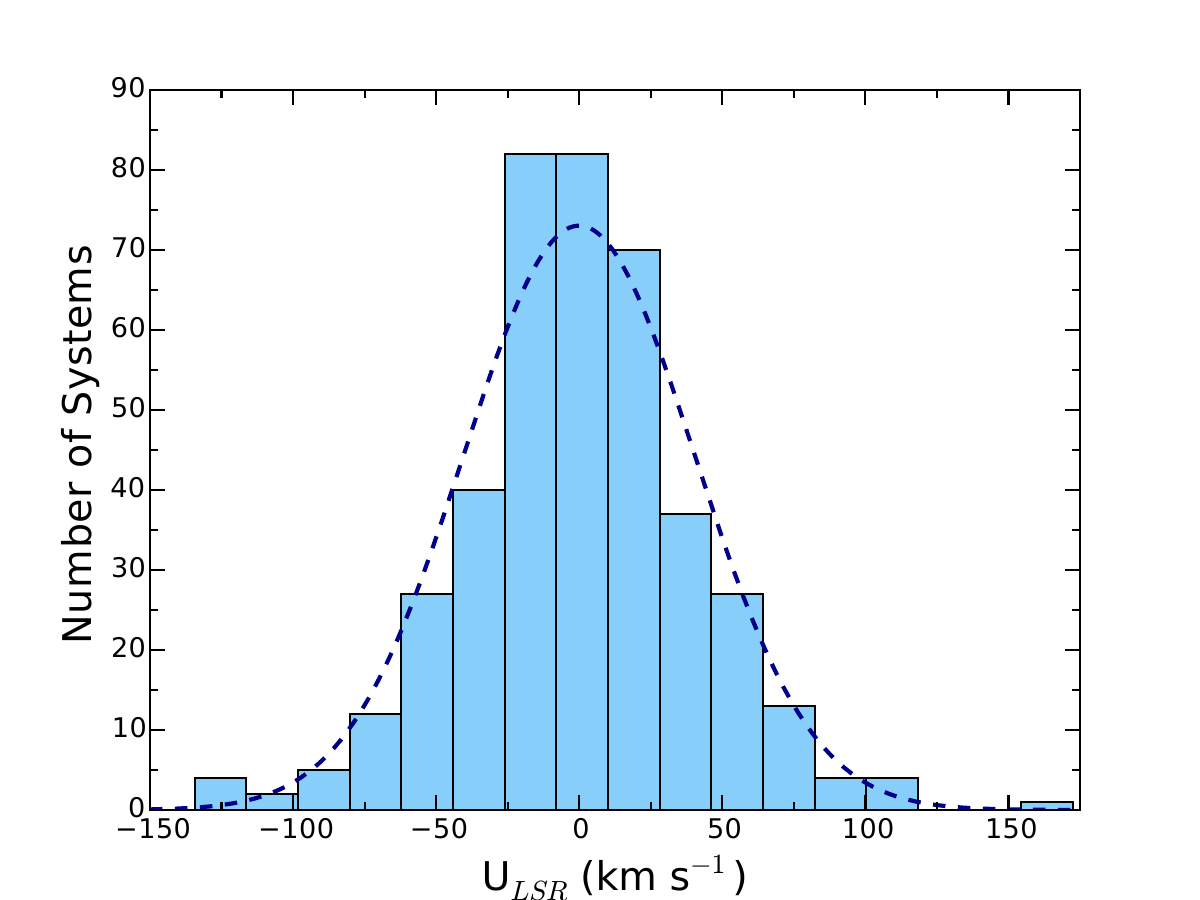}
\hspace{-0.5cm}
\includegraphics[scale=.3,angle=0]{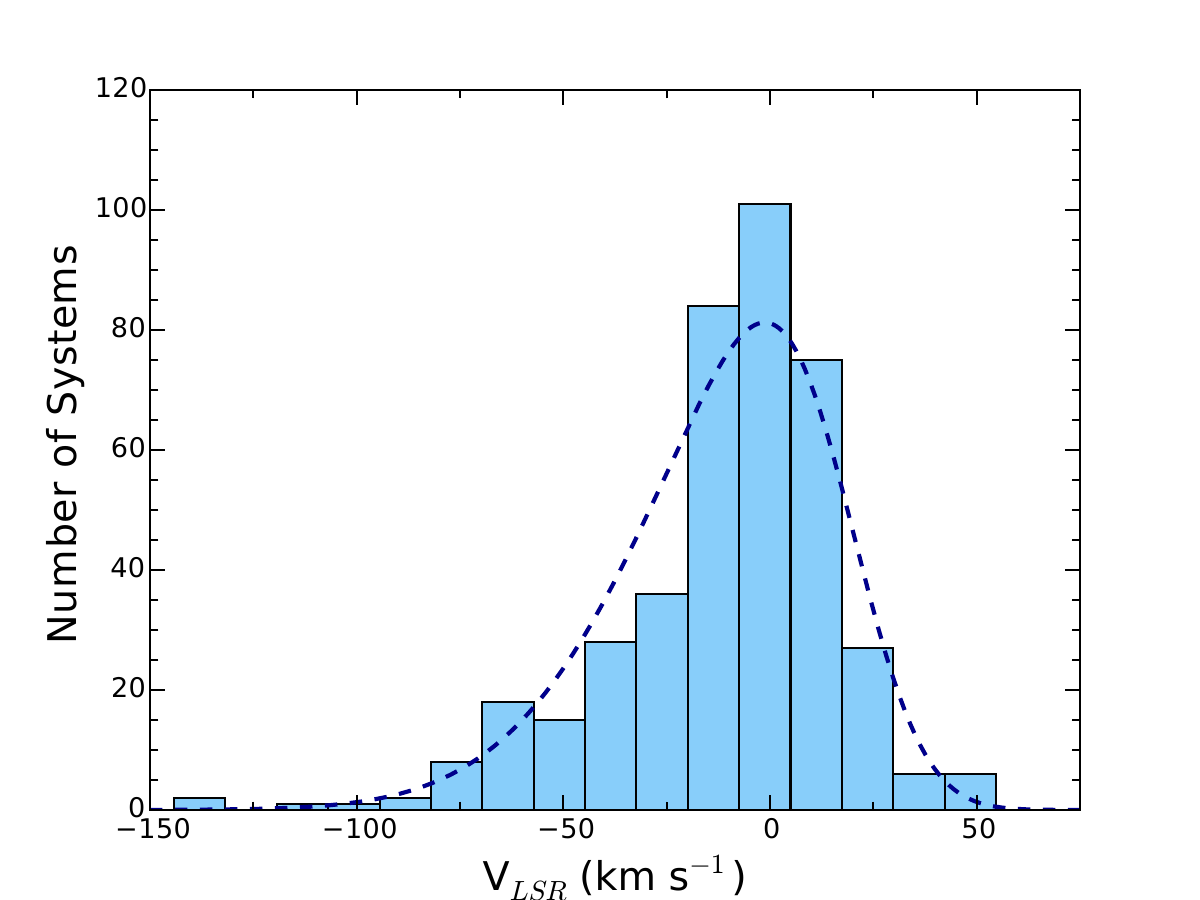}
\hspace{-0.5cm}
\includegraphics[scale=.3,angle=0]{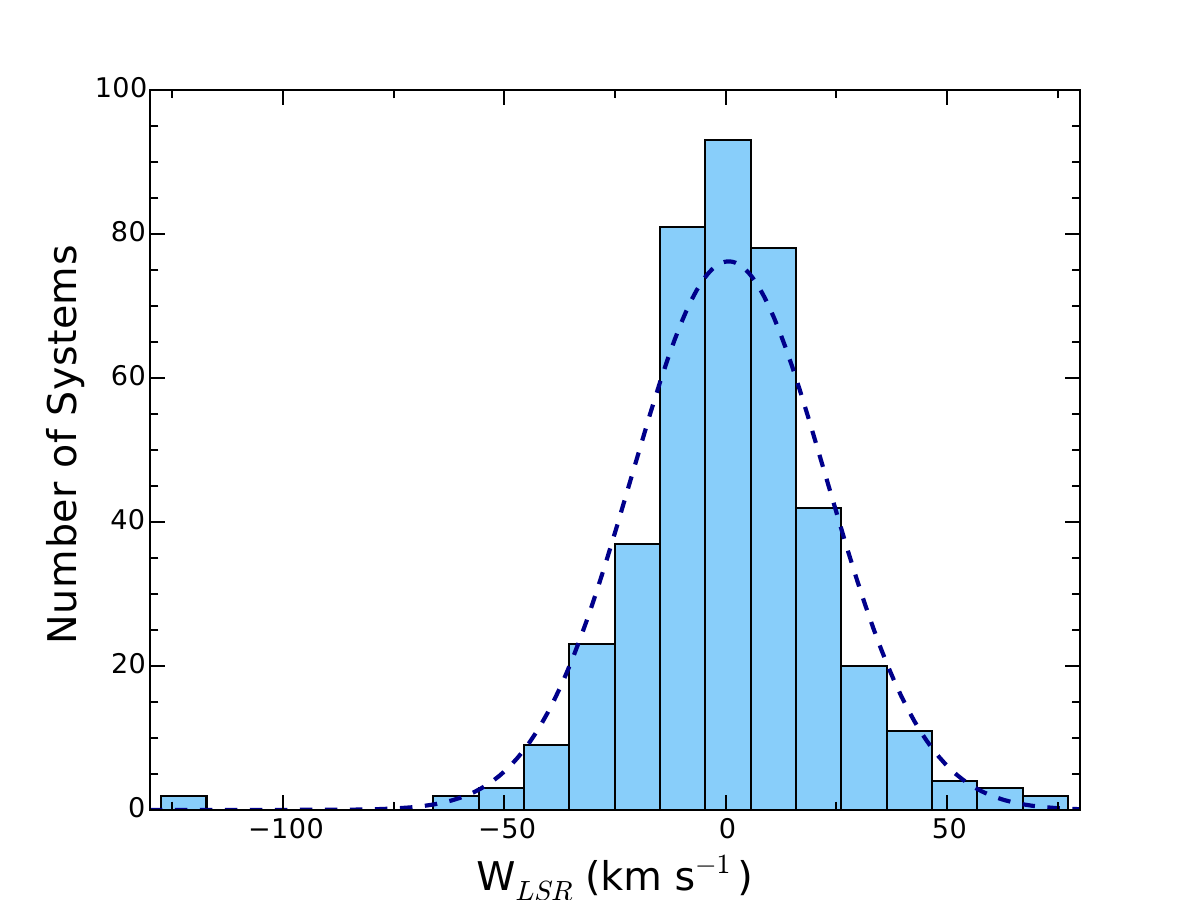}
\caption{The distributions of $UVW$ space motions for our volume-complete sample of 410 mid-M systems within 15 pc, corrected to the local standard of rest (LSR). The distributions of the $U$ and $W$ velocity components are both Gaussian (dashed lines) in shape, while the distribution of the $V$ velocity component is skewed toward negative velocities due to the asymmetric drift of old stars in our sample.   \label{fig:uvw_distr}}
\end{figure*}

In Figure \ref{fig:uvw_distr}, we show the distributions of the ($UVW)_{LSR}$ space velocities for the 410 mid-M dwarf systems in our core spectroscopic sample. The distributions of $U$ and $W$ are generally symmetric, while the distribution of $V$ has a sharp cut-off at roughly 50 km \pers ~and a long tail toward negative velocities. We thus fit Gaussian curves to the $U$ and $W$ velocity distributions and fit a skewed normal curve to the $V$ velocity distribution. The resulting means are (-0.04, -10.8, 0.7) km \pers ~and the standard deviations from the mean are (40.4, 27.4, 22.0) km \pers ~for $UVW$, respectively. For the U and W distributions, the peaks and means are the same due to their symmetric nature. But because the $V$ distribution is skewed, with a skew of -2.7, the peak and the mean differ, where the peak is at -1.2 km \pers ~and the FWHM of the skewed fit is 40 km \pers. This skew is due to asymmetric drift, where stellar populations with larger random motions (older populations) tend to have a mean galactic rotation velocity $V_{LSR}$ that lags behind that of the LSR, as described in \citet{Binney(1998)}. 

We also investigated the $UVW$ distributions when considering only the 366 primary stars in our sample, as well as for 328 effectively single mid-M dwarfs. As shown in Table \ref{tab:uvw}, neither of these distributions differ significantly from that of our core spectroscopic sample.

\newpage

\startlongtable
\begin{deluxetable*}{lcl}
\tabletypesize{\scriptsize}
\tablecaption{Astrometric, Spectroscopic, and Multiplicity Data for 540 Nearby M Dwarfs \label{tab:sample_data}}
\tablehead{\colhead{Column}          &
	   \colhead{Units}               &
	   \colhead{Description}         
	   }
\startdata
1   &  hh:mm:ss  & Right Ascension (J2000.0)                      \\
2   &  dd:mm:ss  & Declination (J2000.0)      			  \\
3   &            & Name    					  \\
4   &            & Component    				  \\
5   &            & 2MASS Identifier   				  \\
6   &            & Configuration of Multiple Components 	  \\
7   &  arcsec    & Separation of Multiple Components   		  \\
8   &            & Reference for Multiplicity Information  	  \\
9   &  mas/yr    & Proper Motion in RA  			  \\
10  &  mas/yr    & Uncertainty in Proper Motion in RA 		  \\
11  &  mas/yr    & Proper Motion in Dec    			  \\
12  &  mas/yr    & Uncertainty in Proper Motion in Dec  	  \\
13  &            & Reference for Proper Motion    		  \\
14  &  mas       & Parallax  					  \\
15  &  mas       & Uncertainty on Parallax 			  \\
16  &            & Note on parallax (1)   			  \\
17  &            & Reference(s) for Parallax  			  \\
18  &  km/s      & Rotational Broadening  			  \\
19  &  km/s      & Rotational Broadening rms  		  \\
20  &            & Note on Rotational Broadening (2) 		  \\
21  &  km/s      & Rotational Broadening from Literature  	  \\
22  &  km/s      & Uncertainty on Literature Rotational Broadening\\
23  &            & Limit for Literature Rotational Broadening (2) \\
24  &            & Reference for Literature Rotational Broadening \\
25  &  km/s      & Weighted Mean Radial Velocity    		  \\
26  &  km/s      & Weighted Mean Radial Velocity Uncertainty 	  \\
27  &  km/s      & Radial Velocity rms 			  \\
28  &            & Number of Measurements 			  \\
29  &            & Instrument 					  \\
30  &  yr        & Baseline 					  \\
31  &  km/s      & Radial or Systemic Velocity from Literature    \\
32  &  km/s      & Uncertainty on Literature Radial or Systemic Velocity \\
33  &            & Reference for Literature Radial or Systemic Velocity \\
34  &  km/s      & Adopted Radial or Systemic Velocity  	  \\
35  &  km/s      & Uncertainty on Adopted Radial or Systemic Velocity \\
36  &            & Note on Adopted Radial or Systemic Velocity (3)\\
37  &  km/s      & U  						  \\
38  &  km/s      & Uncertainty on U 				  \\
39  &  km/s      & V  						  \\
40  &  km/s      & Uncertainty on V  				  \\
41  &  km/s      & W  						  \\
42  &  km/s      & Uncertainty on W  				  \\
43  &            & Galactic Population Membership   		  \\
44  &  Msol      & Stellar Mass 				  \\
45  &  Msol      & Uncertainty on Stellar Mass 			  \\
46  &            & Reference for Stellar Mass 			  \\
47  &  d         & Photometric Rotation Period 			  \\
48  &            & Reference for Photometric Rotation Period 	  \\
49  &            & Note on Sample Membership (4)                  \\
\enddata
\tablecomments{(1) The codes for notes on the DR2 parallax are: 'w' --- weighted mean of two components; 'A' --- adopted parallax and proper motion of primary component; '2p' --- 2 parameter astrometric solution; 'n' --- no DR2 data point.}
\tablecomments{(2) Note on Rotational Broadening: '$<$' indicates an upper limit on the $v \sin i$.}
\tablecomments{(3) Note on Adopted Gamma Velocity: 'a'-- gamma is adopted from mid-M component; `c'-- gamma from composite spectra; `m'-- gamma is weighted mean; `n'-- weighted mean includes gamma from orbit; `o'-- gamma from orbit; `s'--  presumed single; `t' -- gamma calculated as average of component RVs from todcor or tricor; `w'-- gamma from Wilson method.}
\tablecomments{(4) Note on Sample Membership: `m': member of 15 pc mid-M dwarf spectroscopic sample; `n': non-member of 15 pc mid-M dwarf spectroscopic sample. }
\tablerefs{this work;
\citet{Allen(2012)};
\citet{Barnes(2014)};
\citet{Baroch(2018)};
\citet{Bartlett(2017)};
\citet{Benedict(1998)};
\citet{Benedict(2016)};
\citet{Bergfors(2010)};
\citet{Berta(2011)};
\citet{Berta-Thompson(2015)};
\citet{Beuzit(2004)};
\citet{Bowler(2015)};
\citet{Browning(2010)};
\citet{Burningham(2009)};
\citet{Chubak(2012)};
\citet{Cortes-Contreras(2017)};
\citet{Costa(2005)};
\citet{Daemgen(2007)};
\citet{Dahn(1988)};
\citet{Davison(2015)};
\citet{Delfosse(1998)};
\citet{Delfosse(1999c)};
\citet{Delfosse(1999d)};
\citet{Deshpande(2012)};
\citet{Deshpande(2013)};
\citet{Dieterich(2012)};
\citet{DiezAlonso(2019)};
\citet{Dittmann(2014)};
\citet{Dupuy(2017)};
\citet{Engle(2023)};
\citet{Farihi(2005)};
\citet{Finch(2016)};
\citet{Finch(2018)};
\citet{Fouque(2018)};
\citet{GaiaDR3(2023)};
\citet{Gatewood(2008)};
\citet{Gatewood(2009)};
\citet{Gilhool(2018)};
\citet{Gizis(1997b)};
\citet{Gizis(2002)};
\citet{Golimowski(2004b)};
\citet{Harrington(1980)};
\citet{Harrington(1981)};
\citet{Harrington(1985)};
\citet{Hartman(2011)};
\citet{Hawley(1996)};
\citet{Heintz(1974)};
\citet{Heintz(1990)};
\citet{Heintz(1993)};
\citet{Heintz(1994)};
\citet{Henry(1999)};
\citet{Henry(2006)};
\citet{Henry(2018)};
\citet{Hershey(1998)};
\citet{Horch(2011a)};
\citet{Horch(2015)};
\citet{Houdebine(2012a)};
\citet{Houdebine(2015)};
\citet{Houdebine(2016)};
\citet{Ireland(2008)};
\citet{Janson(2012)};
\citet{Janson(2014a)};
\citet{Jao(2003)};
\citet{Jeffers(2018)};
\citet{Jenkins(2009)};
\citet{Jodar(2013)};
\citet{Kesseli(2018)};
\citet{Kiraga(2012)};
\citet{Kohler(2012)};
\citet{Lafarga(2020)};
\citet{Lacy(1977)};
\citet{Lepine(2003)};
\citet{Lepine(2009)};
\citet{Lindegren(1997)};
\citet{Lindegren(2018)};
\citet{Lindegren(2021)};
\citet{Lowrance(2002)};
\citet{Luyten(1979b)};
\citet{Luyten(1997)};
\citet{Malo(2014)};
\citet{Mamajek(2013)};
\citet{Martinache(2009)};
\citet{Mason(2018)};
\citet{Mason(2019)};
\citet{Medina(2020)};
\citet{Medina(2022b)};
\citet{Mohanty(2003)};
\citet{Montagnier(2006)};
\citet{Morales(2009)};
\citet{Newton(2014)};
\citet{Newton(2016)};
\citet{Newton(2018)};
\citet{Nidever(2002)};
\citet{Pass(2022)};
\citet{Pass(2023b)};
\citet{Pravdo(2004)};
\citet{Reid(1995)};
\citet{Reiners(2009)};
\citet{Reiners(2010)};
\citet{Reiners(2012)};
\citet{Reiners(2018)};
\citet{Riedel(2010)};
\citet{Riedel(2011)};
\citet{Salama(2022)};
\citet{Salim(2003)};
\citet{Scholz(2004a)};
\citet{Segransan(2000)};
\citet{Shkolnik(2012)};
\citet{Skrutskie(2006)};
\citet{Smart(2010b)};
\citet{Soderhjelm(1999)};
\citet{Stelzer(2016)};
\citet{SuarezMascareno(2016)};
\citet{Terrien(2015)};
\citet{Tinney(1998)};
\citet{Tokovinin(2012c)};
\citet{Torres(2006)};
\citet{vanAltena(1995)};
\citet{vanLeeuwen(2007)};
\citet{Vrijmoet(2020)};
\citet{Vrijmoet(2022)};
\citet{Ward-Duong(2015)};
\citet{Weinberger(2016)};
\citet{West(2015)};
\citet{Winters(2017)};
\citet{Winters(2018)};
\citet{Winters(2019a)};
\citet{Winters(2019b)};
\citet{Winters(2020)};
\citet{Winters(2022)};
\citet{Woitas(2003)}
.}
\tablecomments{This table is available in its entirety in machine-readable form.}
\end{deluxetable*}

\subsection{Galactic Population Membership}

We can now assign membership to the Galactic populations, based on our $(UVW)_{LSR}$ space motions. We use the prescription in \citet{Fantin(2019)}, based on white dwarf kinematics, to estimate which systems in our sample are most probably thin disk, thick disk, or halo members. 

We use methods described in \citet{Bensby(2003)}, quantified in equations $1 - 3$ in that paper, to calculate each the system's thin disk, thick disk, and halo membership probability $f(U,V,W)$ using the
(U, V, W)$_{LSR}$ space velocity of each star and the typical asymmetric drift and velocity dispersion in each component for each population. We then multiply each probability by the number density of stars present in each
component. Finally, we calculate the ratios of the probabilities, as in \citet{Bensby(2003)}. We adopt as highly-probable members stars with $P(thick)/P(thin) \geq$ 10, which are ten times \textit{more} likely to be members of the thick disk members than of the thin disk (highly probable thick disk members). In contrast, stars with $P(thick)/P(thin) \leq$ 0.1 are ten times \textit{less} likely to be thick disk members than thin disk members, and as a result are highly probable thin disk members. We also calculate $P(thick)/P(halo)$ and assign systems with ratios $\leq$ 0.1 as highly probable halo members.  For convenience, we replicate equations $1 - 3$ from \citet{Bensby(2003)} here:

\small

\begin{equation}
    f(U,V,W) = k \cdot exp \left(- \frac{U^{2}_{LSR}}{2 \sigma^{2}_{U}} - \frac{(V_{LSR} - V_{asym})^2}{2 \sigma^{2}_{V}} - \frac{W^{2}_{LSR}}{2 \sigma^{2}_{W}} \right), 
\end{equation}

\normalsize
where 

\begin{equation}
    k = \frac{1}{(2 \pi)^{3/2} \sigma_U \sigma_V \sigma_W} .
\end{equation}

\small
\begin{equation}
    \frac{P(thick)}{P(thin)} = \frac{X_{thick}}{X_{thin}} \cdot \frac{f(thick)}{f(thin)};  \frac{P(thick)}{P(halo)} = \frac{X_{thick}}{X_{halo}} \cdot \frac{f(thick)}{f(halo)}
\end{equation}

\normalsize
\noindent We use recent ($UVW$)$_{LSR}$ velocity dispersions of (33, 15, 15) km \pers ~for the thin disk, (40, 32, 28) km \pers ~for the thick disk, and (131, 106, 85) km \pers ~for the halo populations from \citet{Fantin(2019)}, assuming X, the fractions of stars in each population, are 83\%, 17\%, and 0.1\% (from \citealt{Bensby(2003)}).

Of our 410 sample members, we determine that 331 systems (81\%) are highly probable thin disk population members and 33 systems (8\%) are highly probable thick disk population members. The  remaining 46 systems (11\%) are neither highly probable thin or thick disk members, and we thus label them with the `ambiguous' population membership assignment. We do not identify any of our sample stars as highly probable members of the halo population. 
We note here that for the nine unresolved binaries with composite spectrum from which we adopted the large RV rms as the RV uncertainty, we tested whether the system's population membership was affected by the large RV uncertainty. We recalculated our $UVW$ space velocities and their respective $P(thick)/P(thin)$ 
ratios using both the upper and lower limits on the RV, and found that all of the systems remained highly probable thin disk members.

\citep{Cortes-Contreras(2024)} recently published a kinematic study of roughly 2200 M dwarfs with spectral types ranging from M0 V to M9 V at distances largely within 50 pc and north of $\delta = -30$\arcdeg. Compared to our membership assignment to only thin or thick disks, the authors of that study divided their sample into more finely defined disks -- young, thin, thick-thin, thick -- and halo populations. If we consider our highly probable thin disk population as similar to the combination of their young and thin disk populations and our thick disk similar to their thick disk, our percentages are generally in agreement: 81\% vs. 89.5\% and 8\% vs. 6.8\%, respectively.

Table \ref{tab:sample_data} lists sample data for 540 M dwarfs. Of these 540, 413 are the mid-to-late M dwarfs in our core spectroscopic sample, while the additional 127 M dwarfs were observed as part of an extended sample. Columns 1-2 note the J2000.0 coordinates, while our preferred name, component, and 2MASS identifier are listed in columns 3-5. Multiplicity information is noted in columns 6-8, while astrometry data are given in columns 9-17. As noted in \citet{Winters(2021)}, we have adopted the Gaia DR2 parallaxes for most of our targets. When DR2 parallaxes were unavailable, we adopted parallaxes from ground-based efforts. Spectroscopic results from our survey are noted in columns 18-20 (rotational broadening) and 25-30 (RVs, instrument, and baseline). We list measurements from the literature in columns 21-24 (rotational broadening) and 31-33 (RVs). In columns 34-36 we note the best or adopted RV and uncertainty used in our Galactic space motion calculations, along with a code identifying the source of those values. Columns 37-43 list the $UVW$ space motions, associated uncertainties, and population membership for the 410 stars in our core spectroscopic sample. Stellar mass information is listed in columns 44-46, while photometric rotation period data are given in columns 47-48. Finally, a sample membership code is noted in column 49.

\subsection{Rotation as a Function of Stellar Mass}

Because rotational broadening measurements from spectroscopic studies and rotation period measurements from photometric studies are two different methods to measure stellar rotation, we can compare our $v \sin i$ measurements and photometric rotation periods from the literature as functions of mass.

\subsubsection{Rotational Broadening by Stellar Mass}

\begin{figure*}
\hspace{-0.3cm}
\includegraphics[scale=.42,angle=0]{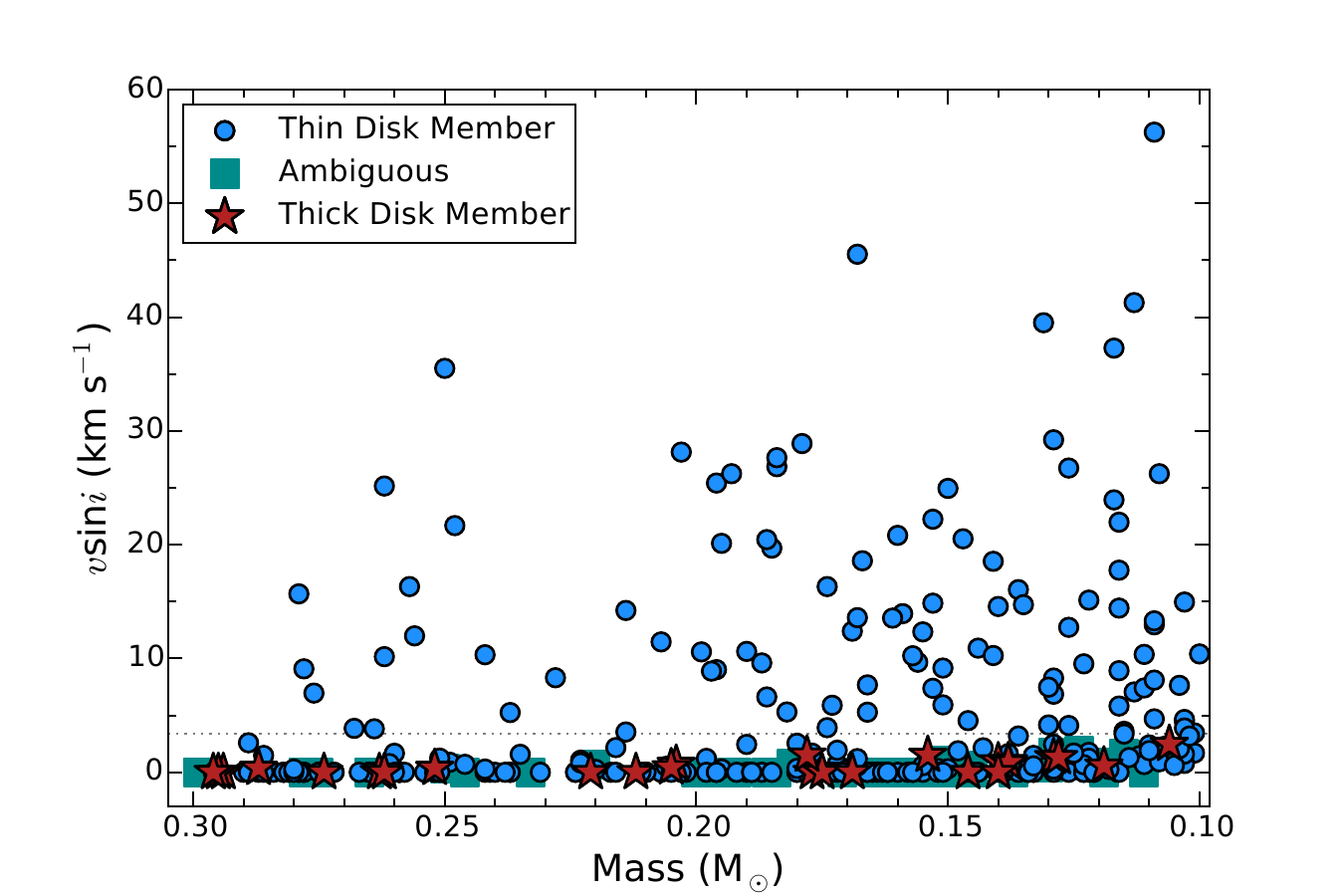}
\hspace{-0.9cm}
\includegraphics[scale=.42,angle=0]{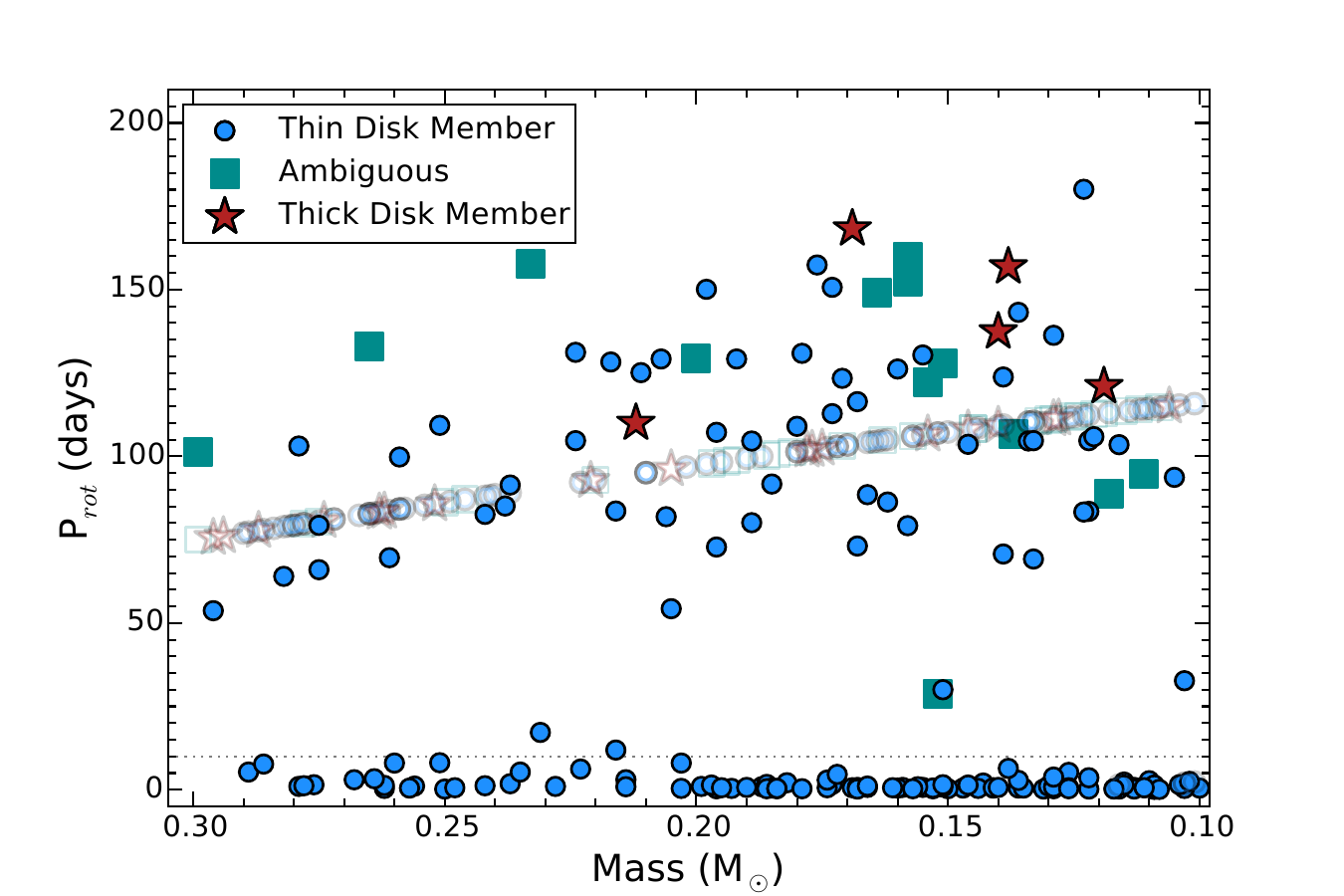}
\caption{Rotational broadening (\textit{left panel}) and photometric rotation period (\textit{right panel}) as a function of stellar mass for the 328 effectively single stars in our sample. The  probable population membership of each star is noted. \textit{Left panel}: The horizontal dotted line indicates our $v \sin i$ detection limit of 3.4 km \pers. We illustrate that all of the highly probable thick disk members have $v \sin i$ values below our detection limit, while all targets with detectable $v \sin i$ values above our detection limit are highly probable thin disk members. Also of note is the range in $v \sin i$ values for the highly probable thin disk members, with a trend of more broadening for lower mass stars. \textit{Right panel}: The horizontal dotted line indicates a 10-day rotation period. The faint, unfilled markers are the 131 inactive stars in our sample with no measured rotation period and for which we have estimated one using the mass-rotation relation for inactive stars from \citet{Newton(2017)} and have uncertainties of $\pm 22$ days. Of note is the known increase in range of rotation periods with decreasing stellar mass.  \label{fig:vsini_prot_mass}}
\end{figure*}

With our rotational broadening measurements in hand, we investigate trends of $v \sin i$ with mass for the 328 effectively single stars in our sample. As noted earlier, we do not include any known multiple systems with angular separations less than 4\arcsec, as the spectra could be broadened due to the composite spectrum. We illustrate in the left panel of Figure \ref{fig:vsini_prot_mass} the relationship between $v\sin i$ and mass for the effectively single targets in our sample, with each target's population membership noted. A few things are noticeable. For one, there is a trend of increasing rotational broadening with decreasing stellar mass. Second, all of the highly probable thick disk members (i.e., relatively older stars) exhibit rotational broadening below our detection limit. In contrast, all 91 systems with rotational broadening above our detection limit of 3.4 km \pers ~are highly probable thin disk members (i.e., relatively younger stars). Overall, we see that the rotational broadening measurements for the fully-convective M dwarfs in our sample are dependent on both mass and relative age of the star. We see that the less massive stars in our sample exhibit more rotational broadening that the more massive stars in our sample. And we see that the younger (thin disk) stars in our sample exhibit more rotational broadening than the older (thick disk) stars in our sample. 

Our results are not surprising and agree with previous results. \citet{Mohanty(2003)} studied a sample of objects with spectral types M4.0 V to L6 V and found both an increase in rotational broadening with increasing spectral type and with decreasing age. \citet{Reiners(2018)} studied a sample of 328 M dwarfs of spectral types M0 - M9 and also found that lower-mass M dwarfs exhibited more rotational broadening than higher-mass M dwarfs. We compare the same phenomenon but manifested as photometric rotation period in the section below.

\subsubsection{Photometric Rotation Period by Mass}

We can perform a similar analysis of photometric rotation period as a function of stellar mass. We first present new and updated rotation period measurements for two systems. 


We report here a new rotation period for LHS~3593 from MEarth-North data. Following previous results reported from the MEarth group \citep{Newton(2016),Newton(2018),Medina(2020),Medina(2022b),Pass(2023b)}, we followed the method described in \citet{Irwin(2011b)} to measure this rotation period. To summarize, we search for sinusoidal modulations in the data  which we then compare to a null hypothesis, as implemented in the $sfit$ module\footnote{\url{https://github.com/mdwarfgeek}}, while taking into account the common mode (i.e., the contamination from water vapor in Earth's atmosphere).

\begin{figure}
\includegraphics[scale=.43,angle=0]{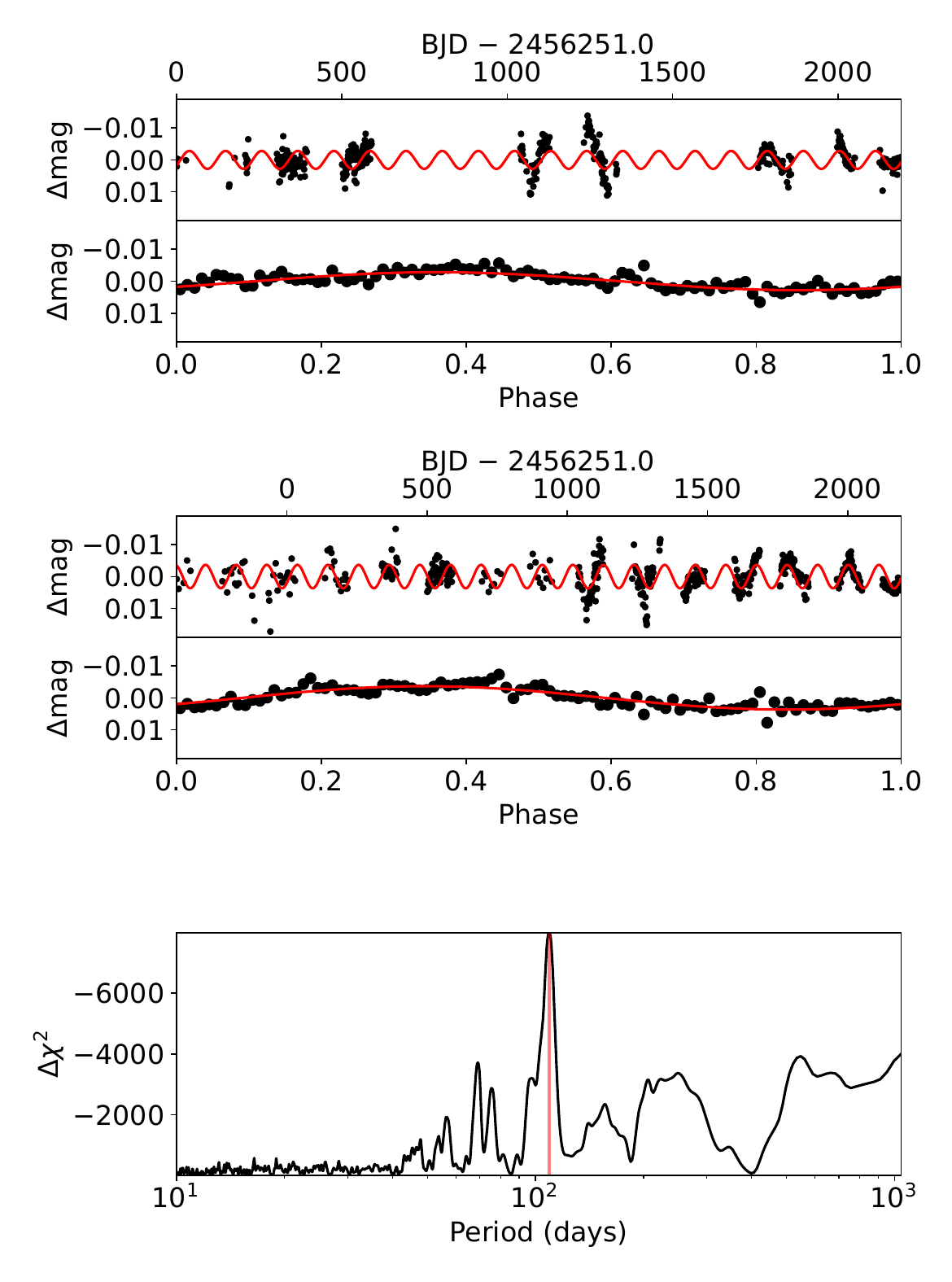}
\caption{Data, fit, phase-folded light curves, and periodogram for the MEarth-North observations of LHS~3593. The top and middle panel set are results from the two separate MEarth-North telescopes used for the observations. The upper of each panel set shows the telescope light curve binned by night, the lower panel shows the light curve binned into 100 phase bins.  The model overplotted corresponds to the alternate hypothesis under test, where the delta magnitude has been corrected for the magnitude zero point and common mode terms, leaving only the modulation. The periodogram in the bottom panel illustrates the change in $\chi^2$ statistic of each alternate hypothesis compared to the null hypothesis of no modulation as a function of period. The best fitting period of 109.0 days is determined by interpolating the $\chi^2$ minimum.   \label{fig:lhs3593}}
\end{figure}

In Figure \ref{fig:lhs3593}, we show two light curves corresponding to the data from two MEarth-North telescopes, analyzed with joint period but separate amplitude, phase, common mode coefficient, and a constant zero point magnitude for each light
curve segment (the ``DC offset" terms, corresponding to instrument changes and sides of the meridian), as described in \citet{Newton(2016)}. The best-fitting period of 109.0 days is illustrated by the peak in the periodogram. Following the quality assessments described in \citet{Newton(2016),Newton(2018)}, we designate this star a `grade A' rotator, an instance where the periodicity is definitively detected and the variability is intrinsic to the star. As seen in Figure \ref{fig:lhs3593}, the semi-amplitude of the variability exhibits significant variation, so we do not report a semi-amplitude measurement from these light curves. 

\begin{figure}[t]
\centering
    \includegraphics[width=\columnwidth]{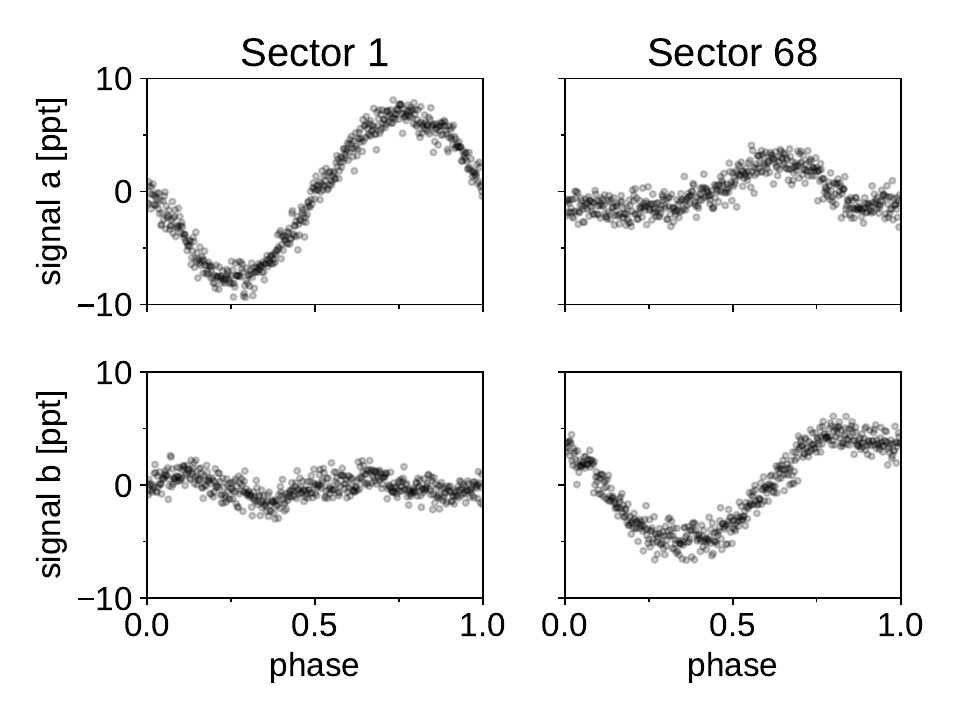}
    \caption{The phased PDCSAP light curves for TIC 161174286 from TESS Sectors 1 (left) and 68 (right), containing blended light from LSPM~J2240-4931AB. The top panels show the 1.002-day signal (with the 1.194-day fit removed) while the bottom panels show the 1.194-day signal (with the 1.002-day fit removed). The points are 500 bins evenly spaced in phase. In Sector 1, the 1.002-day signal dominates the light curve, while the 1.194-day signal is stronger in Sector 68. The first harmonic of the 1.194-day signal is prominent in Sector 1, which is why this harmonic was mistakenly identified as the period in \citet{Pass(2023b)}.}
    \label{fig:LEP2240}
    \end{figure}

We also provide an update to the rotation periods of the wide binary pair LSPM~J2240-4931AB. \citet{Pass(2023b)} identified two rotation periods in the blended light curve of these stars from TESS Sector 1: a strong signal at 1.002 days and a weaker signal at 0.598 days. Since the publication of that work, TESS observed this pair again in Sector 68 \citep{LEP2240_TESS}. The modulation from the second star is much stronger in Sector 68, revealing that the weak 0.598-day signal was actually a harmonic of the true rotation period at 1.194 days (Figure~\ref{fig:LEP2240}). The two stars therefore have rotation periods of 1.002 days and 1.194 days, although it remains unclear which period corresponds to LSPM~J2240-4931A and which corresponds to LSPM~J2240-4931B. For simplicity, we report the marginally longer rotation period as associated with the A component. We emphasize that this has not yet been confirmed with other data.  The addition of the newly measured rotation period for LHS~3593 and for LSPM~J2240-4931A and B brings the total number of our mid-M dwarfs with measured rotation periods to 181.

Of our 328 effectively single sample stars, slightly more than half (181) have published photometric rotation periods. Here, we can assess whether all of the mid-M dwarfs in our sample with measurably broadened spectra have published photometric rotation periods. We see that, when including the new rotation periods for LSPM~J2240-4931A and B, all but four stars with detectable rotational broadening ($>3.4$ km \pers) have published photometric rotation periods. GJ~412B, LSPM~J1805-1422, and LHS~2919 are active in H${\alpha}$ and were discussed in \citet{Pass(2023b)}. These three stars are being monitored by the Tierras Observatory \citep{Garcia-Mejia(2020)} on Mt. Hopkins in Arizona, with newly-measured M-dwarf rotation periods forthcoming (Tamburo, in prep). The remaining star is LHS~1443, an inactive mid-M dwarf that is one of the faintest and lowest-mass targets in our sample ($V = 16.99$ mag, mass $= 0.101$ M$_{\odot}$ ) with $v \sin i$ of $3.46\pm0.61$. Our spectra of this star are extremely low SNR, ranging from 2.5-6.7, making it likely that its rotational broadening probably falls below our detection limit, especially when taking into account the template mismatch discussed in \S \ref{subsubsec:vsini_msmnts}. 

We can estimate the expected photometric rotation periods of our targets from $v \sin i$ and radius using the relation $P_{rot}v \sin i = 2 \pi R \sin i$, where $P_{rot}$ is the photometric rotation period, $R$ is the stellar radius, and $i$ is the inclination to our line of sight of the star's rotation axis. Because of the radius dependence of this relation, there will be a range of expected photometric rotation periods for a given $v \sin i$ for the stars in our sample. For $v \sin i$s greater than 3.4 km \pers, we expect photometric rotation periods $<1.5$ d and $<4.5$ d for mid-M dwarfs with radii $0.1 \rsun$ and  $0.3 \rsun$, respectively. In fact, we see a range in measured rotation periods for the targets with minimally broadened spectra ($v \sin i < 3.4$ km \pers) of $1.4 - 180$ days, while it is $0.1 - 5.2$ days for the targets with detectably broadened spectra ($v \sin i > 3.4$ km \pers). Thus, it is noteworthy that there are 16 effectively single targets with rotational broadening measurements below our detection limit of $v \sin i < 3.4$ km \pers, ~but with measured photometric rotation periods shorter than ten days, periods that should be detectable in TESS data.

We show in the right panel of Figure \ref{fig:vsini_prot_mass} photometric rotation periods as a function of mass for these effectively single mid-M dwarfs with galactic population membership noted. Rotation period measurements are much more challenging for the slowly rotating targets, and thus 140 of the 328 effectively single targets lack measurements. We estimate rotation periods for these H$\alpha$-inactive stars using the relation from \citet{Newton(2017)}  and include them in Figure \ref{fig:vsini_prot_mass} for completeness. We see the same trend as in the left panel of Figure \ref{fig:vsini_prot_mass} where all of the more rapidly rotating targets ($p_{rot} <$ 10 d) are thin disk members, while stars with longer rotation periods belong to both the thin and thick disk populations. We also see a subtle increase in the range of rotation periods with decreasing stellar mass. This is in agreement with results from \citet{Newton(2018)}, as seen in their Figure 4, where a broader range in photometric periods were seen for lower-mass M dwarfs, as compared to higher-mass M dwarfs.

\subsection{$UVW$ Means \& Dispersions of Sample Subsets}

It is well-known that the velocity dispersion of a stellar population in a disk increases with increasing population age \citep{Stromberg(1925),Wielen(1977),Bland-Hawthorn(2016)}. We now have in hand three different measurements that change with a star's age: rotational broadening, photometric rotation period, and galactic disk population membership.  Here we analyze the means and dispersions of our targets' $UVW$ space motions as functions of these measurements to assess any differences between the younger and older sample subsets.

\subsubsection{Complete Sample, Effectively Single Stars, \& Primary Stars}

We fit Gaussian curves and list in Table \ref{tab:uvw} the means and dispersions, along with the number of stars, for 1) our entire volume-complete sample of 410 mid-M dwarfs, 2) the 328 effectively-single mid-M dwarfs, and 3) the 366 systems where the mid-M dwarf is presumed single or is the primary component of a multiple system. A comparison indicates no significant difference between the values for each subsample. The means for $U$ and $W$ are near zero, while the mean in $V$ illustrates a value near $-11$ km \pers, indicative of asymmetric drift. The dispersions in each velocity component are consistent between the three subsamples.

We next perform a comparison between subsamples based on 4) rotational broadening, 5) photometric rotation period, and 6) highly probable thick and thin disk members. The means, dispersions, and number of targets for each group are noted in Table \ref{tab:uvw}.  
\begin{table*}
\begin{longtable*}{@{}lccccccc@{}}
\caption{$UVW$ Velocity Means \& Dispersions of Subsamples \label{tab:uvw}}
\\
\toprule
\multicolumn{1}{c}{Subsample} & \multicolumn{1}{c}{\# stars } & \multicolumn{2}{c}{$U_{LSR}$} &   \multicolumn{2}{c}{$V_{LSR}$}  & \multicolumn{2}{c}{$W_{LSR}$} \\* \cmidrule(lr){1-1} \cmidrule(lr){2-2} \cmidrule(lr){3-4} \cmidrule(lr){5-6} \cmidrule(lr){7-8} 
\endfirsthead
\endhead
\bottomrule
\endfoot
\endlastfoot
     &  &  $\mu$     & $\sigma$     &   $\mu$        &  $\sigma$            & $\mu$     &  $\sigma$   \\*
    &  &  (km s$^{-1}$)     & (km s$^{-1}$)     &   (km s$^{-1}$)        &  (km s$^{-1}$)        & (km s$^{-1}$)     &  (km s$^{-1}$)  \\*
     \cmidrule(lr){3-3} \cmidrule(lr){4-4} \cmidrule(lr){5-5}  \cmidrule(lr){6-6} \cmidrule(lr){7-7}  \cmidrule(lr){8-8} 
Mid-M Dwarfs         & 410 & 0.0  & 40.4   &   -10.8  &  27.4 &  0.7   &  22.0    \\ 
Primaries            & 366 & 0.4  & 39.6    & -11.0 & 27.4 & 0.7   & 21.1      \\
Effectively Single   & 328 & 1.7   & 41.5  & -11.3 & 27.3 & 0.1     & 22.5  \\
\bottomrule
$v \sin i > 3.4$ km \pers & 95 &    $-0.8$  &  $24.8$  & $0.5$  & $12.8$   &  $0.6$   &    $10.8$    \\
$v \sin i < 3.4$ km \pers & 230 &   $2.9$  & $46.5$   & $-15.9$  & $29.9$   &  $0.1$  & $25.8$   \\* 
 \bottomrule
P$_{rot} ~~~~< 10$ d & 107 &    $0.2$  &  $22.6$  & $-0.3$  & $13.5$   &  $0.9$   &    $10.0$    \\
 \bottomrule
Thin Eff. Single & 263 &   $2.6$  &  $37.4$  & $-2.0$  & $17.0$   &  $0.5$   &    $16.0$    \\
 \bottomrule
      &  &  med     & IQR     &   med        &  IQR            & med     &  IQR   \\*
 \bottomrule
P$_{rot} ~~~~< 10$ d & 107 &    $-0.8$  &  $35.2$  & $-1.8$  & $18.4$   &  $0.5$   &    $12.5$    \\
P$_{rot} ~10-90$ d & 26 &   $-12.8$ & $32.3$   & $-6.4$  & $26.3$   &  $1.9$  & $26.8$       \\ 
P$_{rot} ~~~~> 90$ d & 50 & $-13.9$  & $78.7$   & $-10.9$  & $35.9$   &  $-1.9$   & $30.8$       \\* 
 \bottomrule
Thin Eff. Single & 263 &   $1.0$  &  $46.1$  & $-3.3$  & $22.5$   &  $0.2$   &    $20.8$    \\
Thick Eff. Single & 26 &   $-2.1$  & $81.9$   & $-66.9$  & $16.5$  &  $3.9$  & $40.7$          \\* 
\bottomrule
\end{longtable*}
\tablecomments{`Mid-M Dwarfs': our full sample of mid-M dwarfs with RV measurements; `Primaries': presumed single targets and primary components of multiple systems; `Effectively Single': no known stellar, white dwarf, or brown dwarf companion within 4\arcsec.}
\end{table*}

\subsubsection{Rotational Broadening Subsets}
\label{subsubsec:vsini}

We next consider the relation between $UVW_{LSR}$ space velocity and rotational broadening. We choose to consider only the 328 effectively single systems, as the spectral lines may be broadened due to blended spectral lines in the cases of the known multiple systems with composite spectra, rather than rotation. We analyze subsamples for rotational broadening measurement above and below our detection limits. Both subsamples have sufficient numbers of members that their distributions are Gaussian; thus, we fit a Gaussian curve to each subsample, from which we calculate the means and dispersions.

For the 95 effectively single systems in our sample with detectable rotational broadening, all of these targets are highly probable thin disk members and are therefore a kinematically cool group. We see that the means of each velocity component are all near zero, while the dispersions in each velocity component are roughly half that of the entire effectively single subsample.  

\begin{figure*}[t!]
\hspace{-0.2cm}
\includegraphics[scale=.40,angle=0]{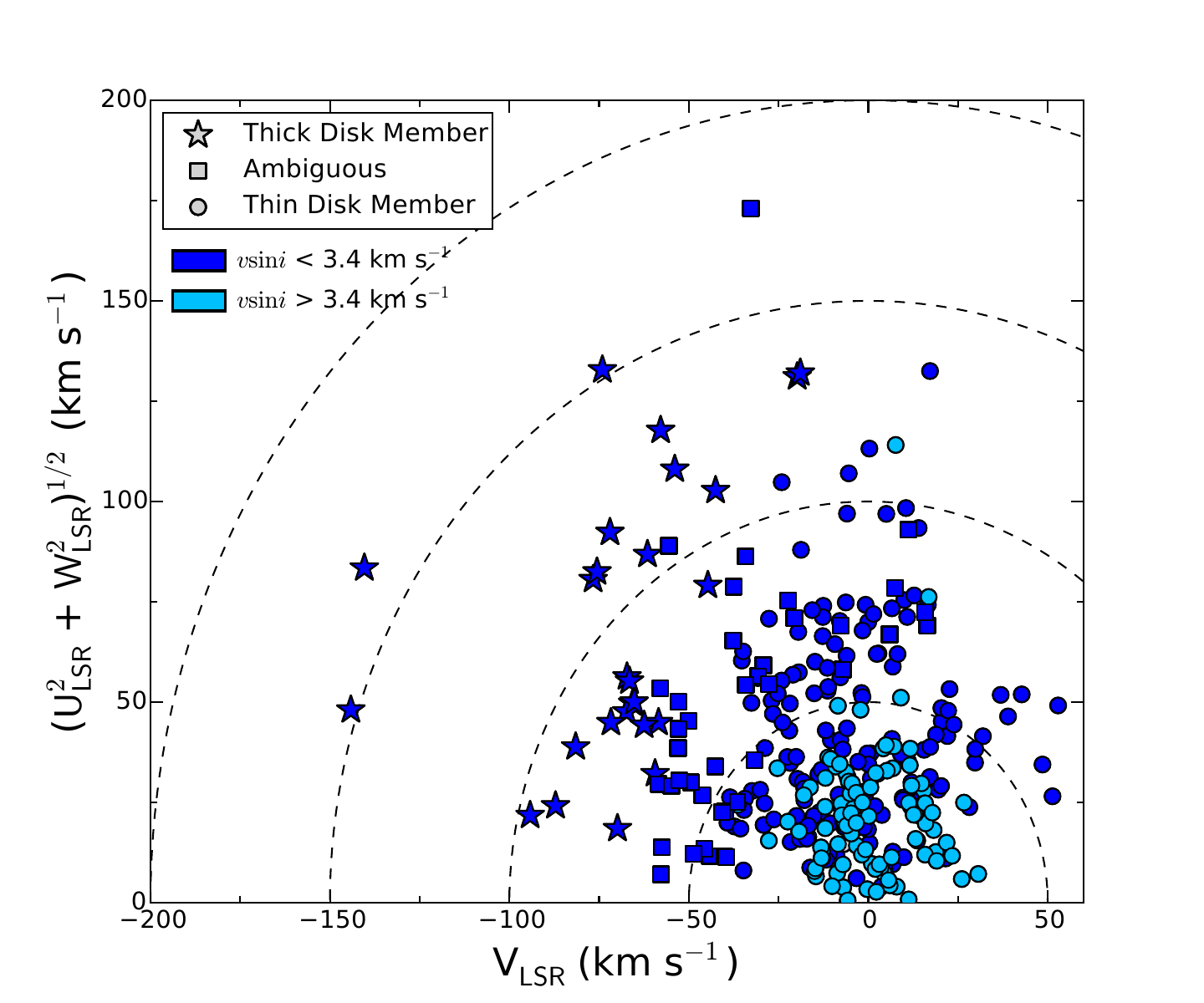}
\hspace{-0.9cm}
\includegraphics[scale=.40,angle=0]{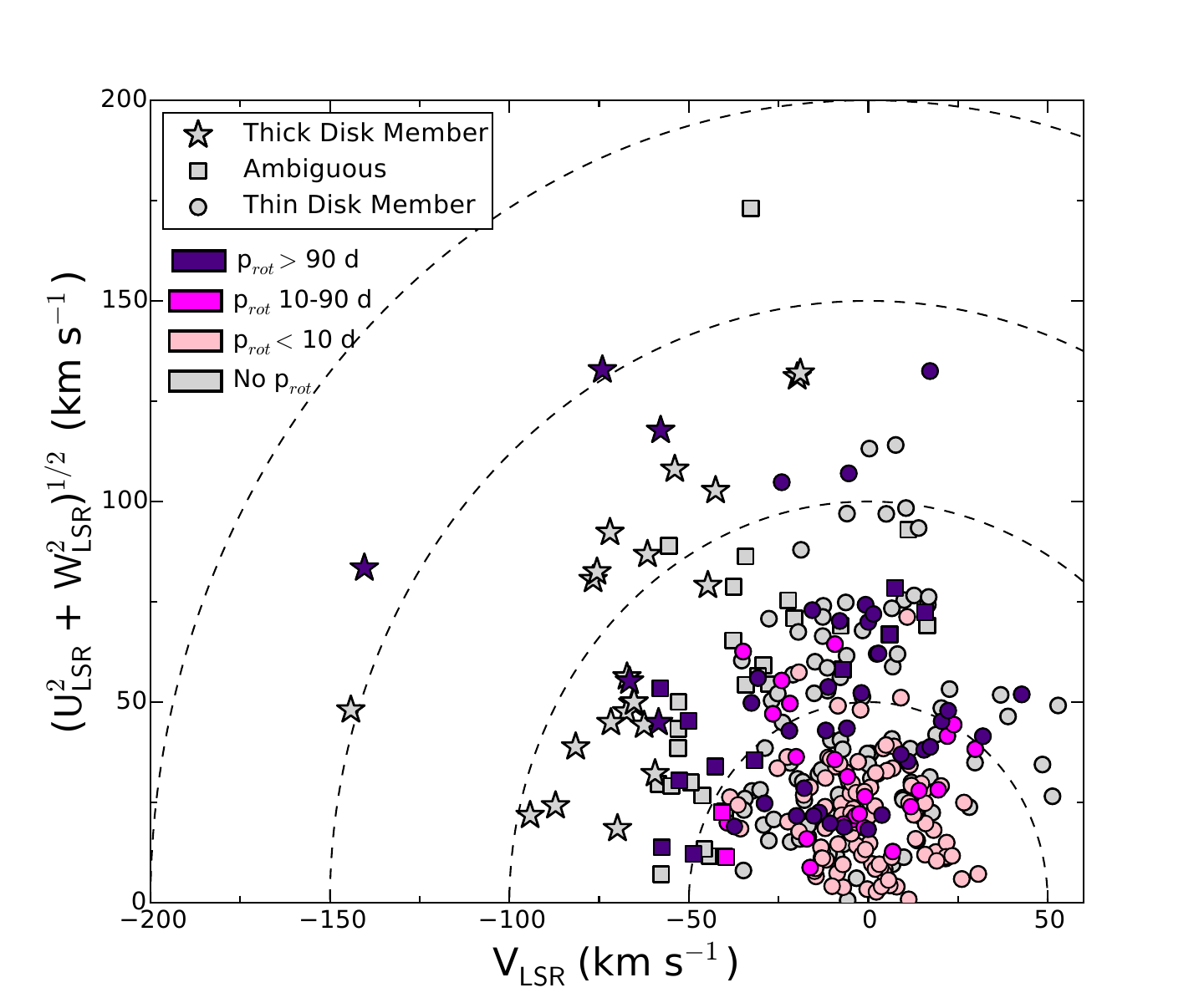}
\caption{Toomre plots of the 328 effectively single mid-M dwarfs in our sample with $v \sin i$ or rotation periods noted. Highly probable thick disk members are shown as stars, highly probable thin disk members are circles, while stars with ambiguous population membership are shown as squares. \textit{Left panel}: We illustrate the rotational broadening measurements for the effectively single mid-M dwarfs in our sample. Those with $v \sin i >$ 3.4 km \pers ~are indicated in pale blue, while those with $v \sin i <$ 3.4 km \pers ~are indicated in dark blue.  \textit{Right panel}: We illustrate photometric rotation periods for the effectively single mid-M dwarfs in our sample that have measurements. Those with $p_{rot} <$ 10 days are indicated in pink, those with $10 < p_{rot} < 90 $ days are indicated in fuchsia, while those with $p_{rot} >$ 90 days are indicated in indigo. Stars with no published rotation period are in gray. \label{fig:toomre_rotations}}
\end{figure*}

For the 230 effectively single stars with $v \sin i <$ 3.4 km \pers, we see that the means of $U$ and $W$ are still near zero, but the mean of $V$ is significantly not zero. This is due to asymmetric drift and can be attributed to the fact that this subgroup of stars is composed of both highly probable thin and highly probable thick disk members, as seen in the left panels of Figures \ref{fig:vsini_prot_mass} and \ref{fig:toomre_rotations}. The dispersions in each velocity component are similar to, if slightly larger than, those of the effectively single subsample, while they are roughly double those of the highly broadened subsample.

We illustrate in the left panel of Figure \ref{fig:toomre_rotations} a Toomre plot with galactic population membership indicated, as well as the rotational broadening measurements for the effectively single mid-M dwarfs in our sample. As expected, all of the stars with detectably broadened spectra are highly probable thin disk members, and all the highly probable thick disk members have $v \sin i$ measurements below our detection limit. But we see that there are also many thin disk members with rotational broadening measurements below that detection limit. Thus, it is not possible to assign galactic population membership for mid-M dwarfs based \textit{only} on rotational broadening if the broadening is less than 3.4 km \pers.  

\subsubsection{Photometric Rotation Period Subsets}

We perform a similar investigation of the relation between the dispersion in galactic space motion component and  photometric rotation period. Of the 328 effectively single stars in our sample, 183 have measured photometric rotation periods. We note that we include in this analysis only the effectively single stars which have published rotation periods; the estimated rotation periods shown in the right panel of Figure \ref{fig:vsini_prot_mass} are not included in our discussion here.

Of these 183 stars, 92 stars with minimally broadened spectra have published rotation periods, while 91 of the effectively single stars with detectable broadening have published rotation periods. For the subsamples with fewer than roughly 100 members, the distributions are distinctly not Gaussian in shape. We therefore calculate the median velocity and inter-quartile range (IQR; the middle 50\% range) for each velocity component, rather than the mean and dispersion, although we supply the results in Table \ref{tab:uvw} from the Gaussian fits for the robust sample of short period rotators (p$_{rot} <$ 10 d). In the discussion that follows, we compare the rotation period subsamples only to each other and not to any other defined subsample.

We divide the effectively single systems into rotation period bins $<$10 day, $10-90$ days, and $>$ 90 days, as in \citet{Medina(2022b)} and \citet{Pass(2023b)}, and calculate the medians and IQRs for each galactic velocity component. We see medians for the $W$ velocity component near zero for all three rotation period subsets, with IQRs in $V$ and $W$ increasing with longer rotation period, as expected due to the increasing relative ages of each subsample. The fact that the $U$ dispersion does not behave similarly we attribute to small number of statistics. We also see an increasingly negative $V$ velocity component median value with increasing rotation period, again in agreement with the expected older population.

Comparison to two recent studies indicates agreement. \citet{Newton(2016)} presented photometric rotation periods, RVs, and $UVW$ space motions for 113 nearby, low-mass M dwarfs (masses $0.1-0.25$ M$_{\odot}$ and trigonometric distances closer than 25 pc). \citet{Medina(2022b)} determined flare rates, photometric rotation periods, H$\alpha$ equivalent widths, and Galactic space motions for a sample of 219 mid-M dwarfs that had been observed by TESS. Both studies report an increase in the spread of $UVW$ velocities, as well as a trend toward increasingly negative $V$ velocity with increasing rotation period.

We illustrate in the right panel of Figure \ref{fig:toomre_rotations} a Toomre plot with galactic population membership and photometric rotation periods indicated for the 183 effectively single mid-M dwarfs in our sample with published measurements. As expected, the targets with the shortest rotation periods ($p_{rot} < 10$ d in pink) are all thin disk members. The five thick disk members with published values all have periods longer than 90 days, but the thin disk members exhibit the full range of rotation periods. Thus, it appears that it is not possible to assign galactic population membership for mid-M dwarfs based \textit{only} on rotation period if the rotation period is longer than 10 days.

\subsubsection{Thin and Thick Disk Subsets}


We also investigate the kinematics of the highly probable thin and thick disk subsets of our sample. The highly probable thin disk subset has a robust number of members, and so we note the results from the Gaussian fit in Table \ref{tab:uvw}. But as with the rotation period subsets, the thick disk subset has few members and does not appear Gaussian in shape. We thus calculated the medians and IQRs of the galactic velocity components for the effectively single stars that were highly probable members of the thin and thick disk populations. We see the same trends as described above for both the rotational broadening and rotation period subsets. In this case, we see a dramatic shift toward more negative values when comparing the median $V$ velocity component of the highly probable thin disk population to the values for the highly probable thick disk members. In general, the dispersion values for the thick disk subset are larger than those of the thin disk, except for the $V$ component. This is likely again due to small number of statistics.

In summary, it appears that effectively single mid-M dwarfs can reliably be assumed to be thin disk members if they exhibit rotation broadening of more than 3.4 km \pers ~and/or if their rotation periods are shorter than 10 days. In contrast, effectively single mid-M dwarfs that are highly probable thick disk members seem to exhibit rotation broadening of less than 3.4 km \pers. However, a star with a measured rotation period of longer than 90 days is not guaranteed to be a thick disk member.

\section{Discussion}
\label{sec:discussion}

\subsection{Stellar and Brown Dwarf Multiplicity Rates}
\label{subsec:mult_rates}

While our sample has not yet been comprehensively surveyed for companions at all separations, we can perform preliminary calculations of the multiplicity rates (MR) of our sample overall, as well as the rates of multiple systems in the thin disk population compared to that of the thick disk population. 

We calculate a stellar MR of $21\pm2$\% for 78 out of 366 systems with stellar companions, which is in agreement with results from other studies. \citet{Winters(2019a)} presented an uncorrected (i.e., minimum) stellar MR of $21.4\pm2.0$\% for a similar mass range ($0.15-0.30$ \msun) of M dwarf primaries. 

This MR is in accord with our understanding that the MR is a decreasing function of primary star mass \citep{Duchene(1999)}. For example, \citet{Raghavan(2010)} demonstrated that 46\% of nearby FGK stars had companions, while \citet{Winters(2019a)} reported an uncorrected MR for early M dwarfs ($0.30 < M_{\odot} < 0.60$) of $28.1 \pm 2.1$\%.

The identification of likely and confirmed brown dwarf companions to LHS~1610 \citep{Winters(2018)}, G~123-45, and LTT~11586 \citep{Winters(2020)} has doubled the number of stars in this sample known to host brown dwarf companions. Five systems in our sample have only brown dwarf companions, while LTT~11586 contains both a stellar and a likely BD companion. This results in an MR for systems hosting brown dwarf companions of $1.6\pm0.7$\% (6/367). Recent work \citep{Fitzmaurice(2023)} has improved our measured minimum mass of $44.8\pm3.2$ M$_{Jup}$ for LHS~1610b to an absolute brown dwarf mass of $50.9\pm0.9$ M$_{Jup}$ via astrometric data from Gaia. Our MR is similar to that for 1120 nearby M dwarfs of all masses within 25 pc (1.3\%; \citealt{Winters(2019a)}) and for nearby solar-type stars ($1.5\pm0.6$\%; \citealt{Raghavan(2010)}). To further compare, \citet{Tokovinin(1992b)} measured an upper limit of $3$\% for the rate of substellar companions (masses $0.02-0.08$ M$_{\odot}$) with periods less than 3000 days around 200 nearby K and M stars more massive than $0.3$ M$_{\odot}$.

\subsection{Multiplicity Rates of Thin and Thick Disk Populations}

\subsubsection{Stellar Multiple Systems}
It has been suggested that all young stars are born in multiple systems \citep{Duchene(2013)}, based on the MRs of pre-main-sequence stars in T associations that were found to be roughly double that of solar-type main-sequence stars \citep{Duchene(1999)}. While we cannot explore in this work whether all mid-M dwarfs are \textit{born} in binary systems because our sample does not contain very many very young stars, we can explore whether there has been any evolution of the MRs of our sample through the lens of their thin (younger stars) and thick (older stars) disk population members. We consider only the 366 systems in our sample that are known to be the primaries of multiple systems or are presumed single. We do not include systems with white dwarf components in these calculations, as the white dwarf was previously the more massive component in the system. To limit ourselves to a consideration of \textit{stellar} multiplicity, we also omit from our calculations systems that have only brown dwarf companions. We also do not include systems with ambiguous population memberships and consider the highly probable members of the thin and thick disk populations only. The total number of primary systems is 366, of which 296 are highly probable thin disk members and 28 are highly probable thick disk members. 

\begin{figure*}
\includegraphics[scale=.75,angle=0]{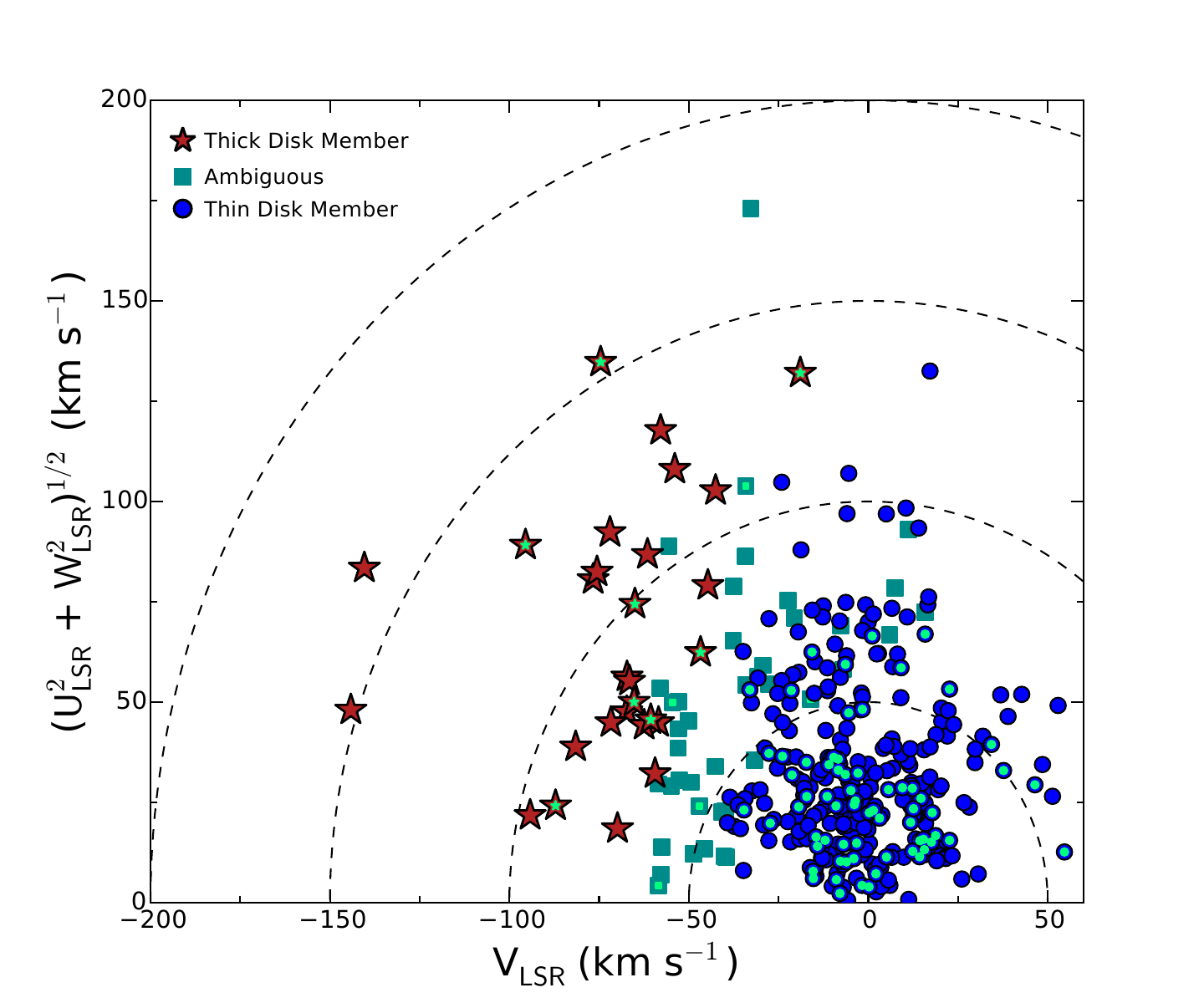}
\caption{Toomre plot of the 366 systems in our sample with mid-M primaries. Highly probable thick disk members are shown as red stars, while the highly probable thin disk members are blue circles. Primaries with ambiguous population membership are shown as teal squares. The primary components of stellar multiple systems (excluding systems that contain a more massive primary, a white dwarf or only a brown dwarf component) are identified in pale green. The dashed lines indicate curves of constant peculiar velocity. The contribution to asymmetric drift from the highly probable thick disk members is evident in the negative $V_{LSR}$ velocities for those members. We find no significant difference between the multiplicity rates of the thin and thick disk samples, when considering the uncertainties.  \label{fig:toomre_binaries}}
\end{figure*}

We show a Toomre plot in Figure 
\ref{fig:toomre_binaries}, with the highly probable thin and thick disk members identified, along with the sample members with ambiguous membership (i.e., those that are \textit{not} highly probable members of either the thin or thick disk populations).  A general trend of increasing velocity dispersion with population is evident, with no clear kinematic boundary between the highly probable thin and thick disk population. The highly probable thin disk members are generally found with $V_{LSR}$ velocities between $\pm 50$ km \pers ~and clustered around $0$ km \pers. The highly probable thick disk members all have markedly  negative $V_{LSR}$ velocities that illustrate the asymmetric drift of old stars.  

Of these, there are 65 and 6 multiple systems in the thin and thick disk populations, respectively. Thus, the stellar multiplicity rates are $22\pm2$\% for the thin disk population and $21\pm8$\% for the thick disk population, where we have used the binomial treatment for the calculations of the uncertainties. This appears to indicate that, at least at the relative ages of our thin and thick disk subsamples, there has been no significant change in the frequency with which mid-M dwarfs host stellar companions. 

Our results are in agreement with conclusions from \citet{Elliott(2014)}, who studied a sample of 192 pre-main-sequence stars in nine loose nearby associations within 150 pc and found no significant difference between the short-period multiplicity rate of their sample when compared to that of field stars and star formation regions.

\subsubsection{White and Brown Dwarf Systems}

We next investigate the population memberships of the systems that contain only brown dwarf companions or a white dwarf component. Of the eight multiple systems with white dwarf components, nearly all are highly probable thin disk members; only GJ~630.1 and GJ~275.2 are highly probable thick disk members.

Of the five multiple systems with only brown dwarf companions, two (GJ~595 and GJ~1001) have a high probability of thick disk membership, two (G~123-45 and LHS~1610) are highly probable thin disk members, while one (GJ~1215) has an ambiguous Galactic population membership. With such small numbers of statistics, we are not able to identify any significant trends in population membership and companion type. There is one system in our sample that is composed of both brown dwarf and stellar components (LTT~11586), but the inclusion of this highly probable thin disk population member does not change our conclusions.

\subsection{Relevance to Exoplanet Studies}

This sample of stars is especially amenable to those searching for planets, as their small masses and radii yield larger planetary RV signals and transit depths than the same planet induces when orbiting a more massive host star. Recent results, based on TESS data, indicate a cumulative occurrence rate of $0.61^{+0.24}_{-0.19}$ for short-period ($0.4-7$ day), terrestrial planets (radii greater than 0.5 R$_{\oplus}$) orbiting mid-to-late M dwarfs \citep{Ment(2023)}. To-date, there are 26 known planetary systems in this sample, of which eight transit their host star. All but two -- GJ~1214  and GJ~12 -- of the eight transiting systems are located in the southern hemisphere. 



We can derive the inclination for stars for which both $v \sin i$ and photometric rotation periods. If planetary orbits are aligned with the host star's rotation axis, this provides a fast way to identify stars that should harbor transiting planets. \citet{Pass(2023a)} analyzed the 90  active stars in our sample that have both rotation periods and rotational broadening measurements above our detection limit and recover the expected isotropic distribution of spin axes.

The multi-epoch radial velocities that we present here provide a resource to help with vetting targets for planetary atmosphere studies. Planetary host stars with specific gamma velocities have been shown to be favorable for detecting molecular oxygen in atmospheres of their planets with the next generation of very high resolution spectrographs, as it is sometimes possible to distinguish the Earth's atmospheric telluric O$_2$ lines from those present in a planet's atmosphere \citep{Lopez-Morales(2019)}.

\subsection{Future Work}

For some of the new multiples discovered as part of this program, we have measured their spectroscopic orbits. These are the subject of a future paper (Winters, in prep). This entire sample of 413 mid-M dwarfs is currently being targeted with high-resolution speckle instruments at the Lowell Discovery 4.3-m and Gemini 8-m telescopes, which will 
complete the coverage at all separation regimes for stellar companions. The results from this speckle program, when combined with our spectroscopic orbits and the Gaia epoch astrometry data expected in future Gaia data releases can be combined to measure absolute masses for these stars. In fact, four binaries in our sample have astrometric orbital solutions in the non-single star catalog from Gaia DR3: LHS~1610, GJ~1029, LHS~1955, and WT~766 \citep{GaiaArenou(2023)}. One of these systems has already been combined with our published RVs to measure the absolute mass of the brown dwarf companion to LHS~1610A \citep{Fitzmaurice(2023)}.  Our complementary program using photometric data from the TESS mission and the MEarth Observatory has resulted in new photometric rotation periods for 80 of our sample stars, the determination of the flare frequency distribution of the single stars in the sample, a dearth of Jovian planet analogs around these stars, spin axis inclinations of our active stars, as well as a relation for the mass-dependent activity lifetime of low-mass mid-M dwarfs \citep{Medina(2020),Medina(2022a),Medina(2022b),Pass(2023a),Pass(2023b),Pass(2024)}. We are also investigating the metallicities of stars in this sample (Pass, in prep). While roughly half of the sample lack measured photometric rotation periods, many of these are ideal targets for such measurements with the newly commissioned Tierras Observatory \citep{Garcia-Mejia(2020)} on Mt. Hopkins in Arizona.

\section{Conclusions}
\label{sec:conclusions}

We have measured multi-epoch rotational and radial velocities for our volume-complete sample of mid-M dwarfs with masses $0.1 - 0.3$ \msun ~that are within 15 pc of the Sun. With these data, we have determined that the majority ($71\pm3$\%) of our volume-complete sample of M dwarfs exhibit minimal rotational broadening  (i.e., $v \sin i$ below our detection limit of 3.4 km \pers).

When combining our RVs with precise astrometric data, we have calculated 3-d space motions for our sample and estimated the probabilities of thin and thick disk membership. We estimated that the majority of our sample (81\%) have a high probability of belonging to the thin disk population, while 8\% have a high probability of being thick disk members. We have demonstrated that the $UW$ space motion distributions are Gaussian in shape, while the distribution of $V$ velocities is skewed due to asymmetric drift, as has been noted by others. These distributions are unaffected when considering only the primary stars in our sample or when including the known mid-M dwarf companions to the primary stars in our sample. 

When considering only the effectively single stars in our sample, we have demonstrated that all of the highly probable thick disk members have $v \sin i$ below our detection limit of 3.4 km \pers. In the same effectively single star sample, we have shown that the stars with $v \sin i$ above our detection limit or with photometric rotation periods less than 10 days are all thin disk members.

We identified seven new multiple systems.  Within 15 pc, LP~69-457 and GJ~376B are two new double-lined binaries, GJ~865 is a new triple-lined system, and GJ~512B exhibits an RV trend that hints at an unresolved companion in a long orbital period. Beyond 15 pc, we have identified RV trends indicative of new companions in long-period orbits to LP~716-10, LHS~5358, and GJ~836. With these new detections, we calculated a preliminary stellar multiplicity rate of $21\pm2$\% and a brown dwarf companion rate of $1.6\pm0.7$\% for mid-to-late M-dwarf primary stars within 15 pc. We also estimated the multiplicity rates of the highly probable thin and thick disk populations in our sample and note that they are statistically the same ($22\pm2$\% for the thin disk population and $21\pm8$\% for the thick disk population), when considering the uncertainties. This seems to indicate minimal change in the frequency of stellar companions between the relative ages of the thin and thick disk populations of mid-M dwarfs considered in our study.

The results from our survey more than triple the number of these nearby, fully-convective stars with complete astrometric data and uniformly derived, multi-epoch, high-resolution RVs and rotational broadening measurements.

\facilities{{FLWO:1.5m (TRES)}, {CTIO:1.5m (CHIRON)}, MEarth, TESS}

\software{IDL, IRAF, {\texttt{python}}}

\vspace{5mm}
\begin{center}
\large
Acknowledgments
\end{center}
\normalsize

We thank the referee for a thoughtful review, which improved the manuscript. We are grateful to members of the TRES team for their help throughout this project, including Allyson Bieryla, Lars Buchhave, Sam Quinn, Andy Szentgyorgyi, and Pascal Fortin. We thank the members of the SMARTS Consortium, who have enabled the operations of the small telescopes at CTIO since 2003. We are indebted to the observers and observer support at CTIO, specifically Arturo Gomez, Mauricio Rojas, Hernan Tirado, Joselino Vasquez, Alberto Miranda, and Edgardo Cosgrove. We are particularly grateful to Todd Henry, Wei-Chun Jao, Hodari-Sadiki James, Leonardo Paredes for CHIRON support, and to Andrei Tokovinin and Debra Fischer for their work on CHIRON. 

This work is made possible by a grant from the John Templeton Foundation. The opinions expressed in this publication are those of the authors and do not necessarily reflect the views of the John Templeton Foundation. This material is based upon work supported by the National Science Foundation under grant AST-1616624, and work supported by the National Aeronautics and Space Administration under Grant No. 80NSSC18K0476 issued through the XRP Program. For a portion of this project, E.P.\ was supported by a Natural Sciences and Engineering Research Council of Canada (NSERC) Postgraduate Scholarship and is currently supported by a Juan Carlos Torres Postdoctoral Fellowship at the Massachusetts Institute of Technology.

This work has made use of data from the European Space Agency (ESA) mission
{\it Gaia} (\url{https://www.cosmos.esa.int/gaia}), processed by the {\it Gaia}
Data Processing and Analysis Consortium (DPAC,
\url{https://www.cosmos.esa.int/web/gaia/dpac/consortium}). Funding for the DPAC
has been provided by national institutions, in particular the institutions
participating in the {\it Gaia} Multilateral Agreement. This research has made use of the Washington Double Star Catalog maintained at the U.S. Naval Observatory, as well as the SIMBAD database and the Aladin and Vizier interfaces, operated at CDS, Strasbourg, France. This work has made use of the Smithsonian Astrophysical Observatory/NASA
Astrophysics Data System. 

\clearpage

\bibliographystyle{aasjournal}
\bibliography{masterref.bib}

\clearpage

\end{document}